\documentclass[%
preprint,  
superscriptaddress,
amsmath,amssymb,
aps,
floatfix,
]{revtex4-1}

\pdfoutput=1
\usepackage{graphicx}% Include figure files
\usepackage{dcolumn}% Align table columns on decimal point
\usepackage{bm}% bold math
\usepackage[usenames,dvipsnames]{xcolor} % mk added
\usepackage[colorlinks,citecolor=MidnightBlue,linkcolor=Orange,urlcolor=MidnightBlue]{hyperref}% add hypertext capabilities
\usepackage[mathlines]{lineno}% Enable numbering of text and display math
\linenumbers\relax % Commence numbering lines
\usepackage{color}
\usepackage{MnSymbol}
\usepackage{times} % mk added
\definecolor{orange}{rgb}{1.0,0.4,0.0}
\definecolor{rose}{rgb}{1.0, 0.33, 0.64}
\usepackage{tikz}
\usepackage{times} % mk added
\usepackage{rotating}
\usepackage{gensymb}
\usepackage{siunitx}
\usepackage{xr}
\usepackage{commath}
\usepackage{hyperref}
\usepackage{cleveref}
\usepackage[draft]{changes}
\usepackage{comment}
\definechangesauthor[name=<Alex>, color=red]{AB}
\usepackage[normalem]{ulem}

\usepackage{esint}

\begin{document}
\nolinenumbers

\title{Phase-separation dependent active motion of Janus lipid vesicles}

\author{V. Willems}
\affiliation{Univ. Bordeaux, CNRS, CRPP, UMR 5031, F-33600 Pessac, France}
\author{A. Baron}%
\email{alexandre.baron@u-bordeaux.fr}
\affiliation{Univ. Bordeaux, CNRS, CRPP, UMR 5031, F-33600 Pessac, France}
\affiliation{Institut Universitaire de France, 1 rue Descartes, 75231 Paris Cedex 05, France}
\author{D. A. Matoz-Fernandez}%
\affiliation{Department of Theoretical Physics, Complutense University of Madrid, Madrid, 28040, Spain}
\author{G. Wolfisberg}
\affiliation{Laboratory of Soft Living Materials, Department of Materials, ETH Zurich, 8093 Zurich, Switzerland}%
\author{E. R. Dufresne}
\affiliation{Laboratory of Soft Living Materials, Department of Materials, ETH Zurich, 8093 Zurich, Switzerland}%
\author{J. C. Baret}%
\author{L. Alvarez}
\email{laura.alvarez-frances@u-bordeaux.fr}
\affiliation{Univ. Bordeaux, CNRS, CRPP, UMR 5031, F-33600 Pessac, France}%

\maketitle

\section*{Abstract}

Active colloidal systems have emerged as promising contenders for the future of microdevices. While conventional designs have extensively exploited the use of hard colloids, the advancement of cell-inspired architectures represents a pivotal path towards realizing self-regulating and highly functional artificial microswimmers. In this work, we fabricate and actuate Janus lipid vesicles demonstrating reconfigurable motion under an AC electric field. The giant unilamellar vesicles (GUVs) undergo spontaneous phase separation at room temperature leading to Janus-like GUVs with two distinct lipid phases. We report self-propulsion of the Janus GUVs via induced charge electroosmosis, in between parallel electrodes. Remarkably, the fluid nature of the lipid membrane affected by the electric field leads to asymmetry-symmetry transient states resulting in run-and-tumble events supported by structure domain analysis. We characterise an enhanced rotational diffusivity  associated with tumble events, decoupled from thermal reorientation. Lastly, we identify cargo-release capabilities and a variety of shape-encoded dynamic modes in these vesicles. This cell-inspired architecture provides an alternative route for creating motile artificial cells and programmable microswimmers.

\section*{Introduction}

Cells, even in their simplest forms, have evolved to execute adaptive motion and basic tasks, a capability encoded in their spatially organized and coherent architecture. Replicating such intricate behaviour at the microscale could pave the way for the next generation of bioinspired microrobots, with significant implications for biomedical devices \cite{Ebbens2016}. 

Active colloids present a suitable minimal model to reproduce the motility of their biological counterparts at the microscale  \cite{Purcell1977,Ebbens2016,Shields2017}, with great potential for the realization of bioinspired microdevices \cite{Gompper2020}. These artificial microswimmers are designed to capture the dynamical features of their biological counterparts such as adaptive motion via external \cite{Colabrese2017,Karani2019, Sprenger2019,Fernandez2020} or internal \cite{Brook2018, Alvarez2021, vanKesteren2023printing} feedback mechanics incorporating responsive materials, or to perform simple tasks \cite{Medina2015, Demirors2018, Aubret2021}. Nevertheless, the use of hard building blocks limits the architecture design and narrows their programmability, in stark contrast to the sophisticated cellular compartments. The realization of hierarchical and compartmentalized designs holds the key to realising cell-mimetic microdevices \cite{Boudet2021}.

A promising alternative is using soft compartments such as giant unilamellar vesicles (GUVs) as biological scaffolds \cite{Walde2010, Rideau2019,Xu2022}. Thanks to their physical-chemical properties, GUVs serve as shape-shifting and semipermeable containers, able to interact with the environment, exchange information, and reconfigure in the presence of obstacles \cite{Walde2010}. Recently, these cell-like architectures have gained an increasing interest in the active matter community as an alternative architecture to construct active cell replicas. First numerical models proposed active giant vesicles driven by internal active colloids as motors  \cite{Paoluzzi2016, Abaurrea-Velasco2019,Peng2022}. While several experimental systems tried to unveil the correlation between membrane deformation and activity of encapsulated active colloids \cite{Vutukuri2020,Sharma2021,Park2022,Song2022}; only recent work achieved motion transfer of encapsulated bacteria in the GUV to the whole membrane \cite{LeNagard2022}. The search for the efficient directed motion of GUVs led to adopting standard propulsion mechanisms found in hard active colloids, such as light- \cite{Bartelt2018} and chemically-driven Janus GUVs \cite{Joseph2017, Somasundar2019, Cui2022, Peng2022}, or protein-mediated GUV transport \cite {Fu2023}. The exciting recent progress on self-propelling giant vesicles hints at the great potential of GUVs as an alternative architecture to design motile artificial cells. Yet, designing adaptive and functional motility such as tactic dynamics reminiscent of biological microswimmers, requires careful design and understanding of the role of lipid membrane properties.

In this context, we introduce an experimental approach to fabricate active cell-mimetic compartments using phase-separated Janus lipid vesicles, with reconfigurable motion. Their rich dynamics under external actuation are determined by the intrinsic membrane fluidity properties, geometry and interaction with their environment. 

\section*{Results}

\subsection*{Janus giant unilamellar vesicles}

\begin{figure*}[h]
\includegraphics[width=\linewidth]{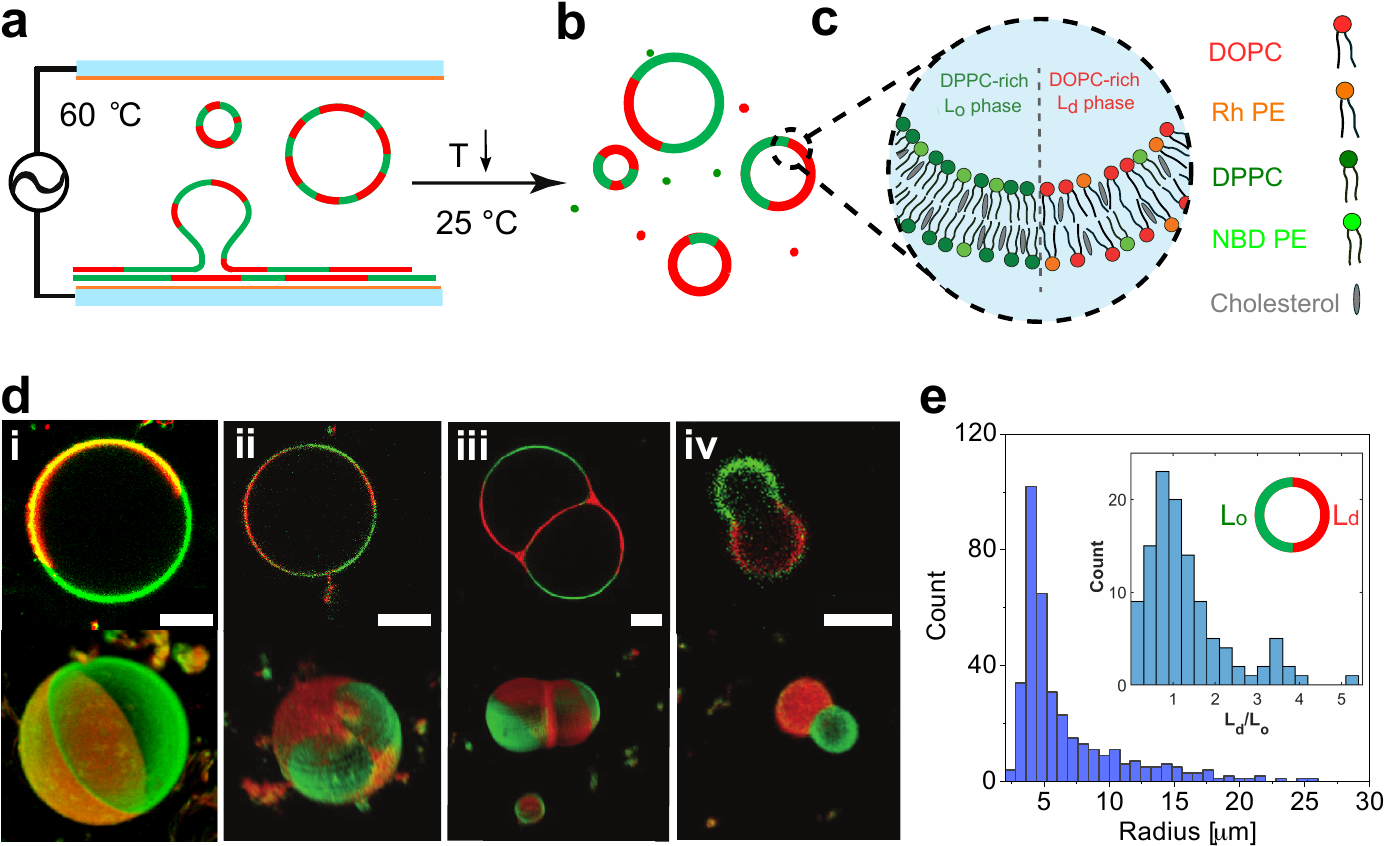}
\caption{\textbf{Fabrication and characterization of Janus phase-separated GUVs}. 
.~\textbf{a,} Scheme of the electroformation process where the dried layers of the lipid mixture are hydrated in water at 60 \degree C in between two electrodes while applying an AC electric field. ~\textbf{b,} Upon cooling to 25 \degree C, spontaneous phase separation arises leading to liquid-liquid phase-separated GUVs with two lipid domains (red and green). \textbf{c,} Scheme of the composition of each phase of the Janus GUV with a liquid-ordered ($\rm L_o$) DPPC-rich domain containing NBDPE (green) and a liquid-disordered ($\rm L_d$) DOPC-rich domain with both NBDPE and RhPE (red). ~\textbf{d,} Confocal and fluorescent microscopy picture of the various geometries obtained i) Janus, ii) patchy, iii) dumbbell, and iv) peanut. The scale bars depict 5 $\rm \mu m$.~\textbf{e,} Size distribution of the spherical Janus GUVs, and asymmetry (inset) distribution as the ratio of the liquid-disordered ($\rm L_d$, red) and the liquid-ordered ($\rm L_o$, green) phases obtained from electroformation.} 
\label{fig:fig1}
\end{figure*}

In this work, we fabricate active and self-reconfigurable phase-separated giant unilamellar vesicles (GUVs), here referred to as Janus GUVs. The Janus GUVs were produced using electroformation \cite{Angelova1986,Drabik2018, Spanke2022} as described in detail in the Methods section, and shown in Fig.\ref{fig:fig1}a. Briefly, using an AC electric field, we hydrate in a solution of 25 mM of sucrose at 60 \degree C a previously dried layer of a ternary lipid mixture: 1,2-dioleoyl-sn-glycerol-3-phosphocholine (DOPC), 1,2-dipalmitoyl-sn-glycerol-3-phosphocholine (DPPC), and cholesterol (Chol), leading to compositionally symmetric GUVs \cite{Veatch2003,Heberle2011} (Supplementary Fig.S1). While cholesterol is miscible in both lipids at a broad range of temperatures, at the chosen composition DOPC and DPPC exhibit spontaneous phase separation below 30 \degree C, forming DOPC-rich liquid disordered ($\rm L_d$) and DPPC-rich liquid-ordered ($\rm L_o$) phases due to the phospholipids unsaturated or saturated nature, with the latter allowing higher packing at lower temperatures \cite{Gudheti2007, guvbook_Ch1}. 

Thus, upon cooling to 25 \degree C, the vesicles exhibited phase separation due to the immiscibility of these two phospholipids. The resulting phase-separated Janus GUVs also contain a small amount of two charged fluorescently labeled phospholipids (see Methods for more details): 1,2-dioleoyl-sn-glycerol-3-phosphoethanolamine-N-(lissamine rhodamine B sulfonyl) (RhPE, red), and 1,2-dipalmitoyl-sn-glycerol-3-phosphoethanolamine-N-(7-nitro-2-1,3-benzoxadiazol-4-yl) (NBDPE, green). NBDPE dissolves in both phases, while RhPE partitions only into the DOPC-rich phase ($\rm L_d$) \cite{Veatch2003}. The presence of the fluorescent phospholipids allows imaging via fluorescence and confocal microscopy to characterize the final GUV population. We image the samples one day after electroformation to ensure the coalescence of the different domains and the formation of Janus-like GUVs \cite{Gudheti2007, Veatch2003}. We characterize mainly 4 types of vesicle conformations : Janus-like, patchy, dumbbells, and peanuts (Fig. \ref{fig:fig1}d, Supplementary Fig.S2). The variety in GUV architecture is related to the low control of the membrane composition with this technique (Supplementary Fig.S2, Supplementary Video 1). 

In this work, we will focus on the study of Janus spherical GUVs (Fig.\ref{fig:fig1}$\rm d_i$), and we will take advantage of their phase separation asymmetry (ratio between $\rm L_d/L_o$ phases) and size polydispersity (\rm 4-6 $\rm \mu m$ radius) to study their rich active dynamics. We characterize the passive dynamics in 2D by image analysis and particle tracking algorithms (see Methods). For the Janus spherical GUVs with a dominant population of $\approx 5$ $\rm \mu m$ radius, we obtained a diffusion coefficient $\rm D_{T} = 0.042~\mu m^2s^{-1}$, consistent with the predicted theoretical value for a particle of the same size (Supplementary Fig.S3).

\subsection*{Giant unilamellar vesicles under perpendicular AC electric field}

\begin{figure*}[h]
\includegraphics[width=0.6\linewidth]{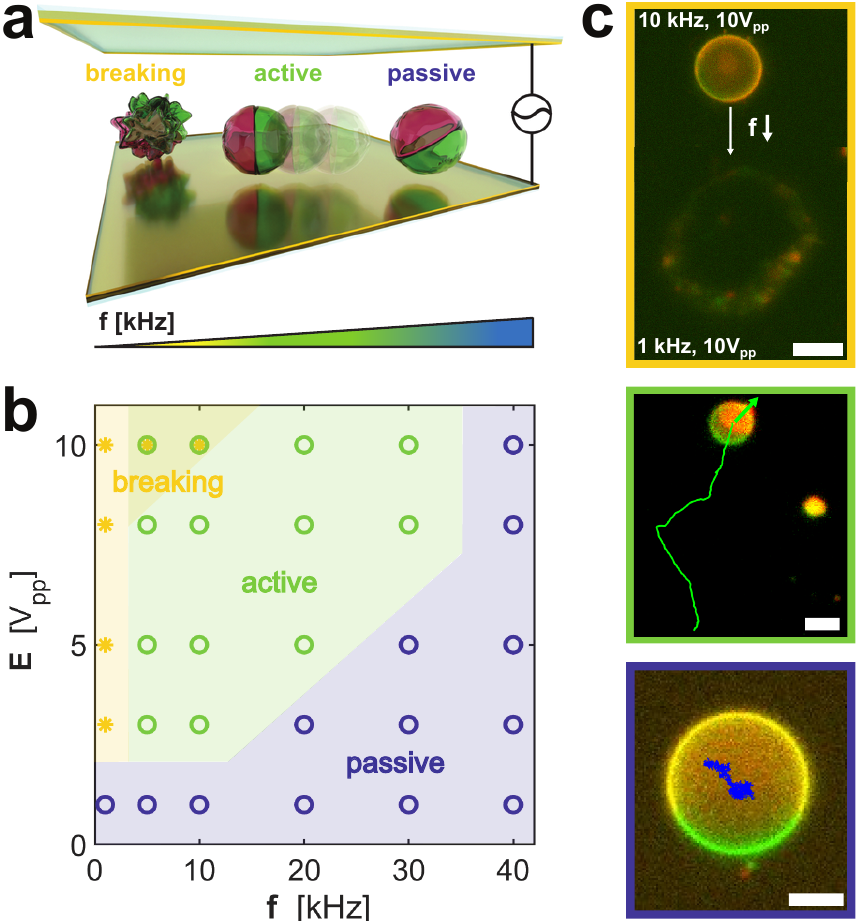}
\caption{\textbf{Behavior of Janus GUVs under AC electric fields} 
.~\textbf{a,} Scheme of the experimental cell composed of two conductive parallel electrodes applying a perpendicular AC electric field, with the corresponding behaviour of the Janus GUVs at varying frequency for passive(blue, PBP), active motion (green, ABP), and bursting (orange, B).~\textbf{b,} Phase diagram of the Janus GUVs at varying field amplitudes $\rm E$ and frequencies $\rm f$ [kHz] of the electric field applied.~\textbf{c,} Fluorescence optical microscopy images depicting the behaviour of the GUVs: bursting (top), active motion (middle), and passive motion (bottom). Scale bars depict 15 $\rm \mu m$, 10 $\rm \mu m$ and 5 $\rm \mu m$ respectively}
\label{fig:fig2}
\end{figure*}

Extensive work on the behaviour of GUVs under electric fields demonstrated the rich behaviour of giant lipid vesicles under AC and DC electric fields in coplanar electrodes, such as membrane deformation,  mixing, or fusion \cite{Dimova2009, Steinkühler2018}. In contrast to the aforementioned examples, our experimental system relies on using parallel electrodes applying a perpendicular AC electric field \cite{Ristenpart2007}. This specific experimental design is based on the actuation mechanism of active Janus hard spheres, with the premise that a spherical Janus GUV should exhibit active dynamics reminiscent of their colloidal analogues. To demonstrate this hypothesis, we study the behaviour of Janus spherical GUVs in this configuration (Fig.\ref{fig:fig2}a). Briefly, we placed 7.4 $\rm \mu L$ of the phase-separated GUV suspension in an experimental cell consisting of a bottom and top transparent ITO-coated electrode connected to a function generator and separated by a 120 $\rm \mu$m spacer. We functionalize the ITO with a negative polyelectrolyte (PSS), causing electrostatic repulsion between the vesicles and the electrode to avoid sticking of the GUVs on the substrate (more details in Methods).

We applied an AC electric field in the low kHz range to probe the behaviour and response of the Janus GUVs. In all cases, the AC field was applied starting from 1 V$_{pp}$ up to 10 V$_{pp}$ (833.3 - 8333.3 V/cm), at 40 kHz to allow equilibration of the cell. This enables the vesicles, otherwise neutrally buoyant due to the density match among the inner and outer solutions, to get attracted to the electrodes due to electrostatic interactions. After equilibration, we probed the entire range of field conditions from 1 to 10 V$_{pp}$ and 40 to 1 kHz via optical fluorescence and confocal microscopy (see Methods). We constructed a phase diagram with the behaviour of Janus vesicles of all sizes at different amplitudes and frequencies of the applied AC field, as shown in Fig.\ref{fig:fig2}b. We observed three different regions: i) vesicle bursting in the narrow yellow region at low frequencies and mid-high amplitudes, ii) active motion in the green region at high amplitudes and medium frequencies, and iii) no active motion depicted in the blue region at high frequencies and low amplitudes.

At low frequencies (1-5 kHz), vesicles burst at field strengths larger than 3 V$_{pp}$. This phenomenon is produced by the accumulation of charges at both sides of the membrane, causing high transmembrane potential and inducing effective electric tension leading to membrane breakage (Supplementary Movie 2) \cite{Riske2009, Dimova2009}. The bursting of vesicles described in literature often occurs under short DC-field pulses (200 $\mu$s) \cite{Dimova2009}, equivalent and consistent to a period of an AC field with a 5 kHz frequency, where we start to observe bursting events.

At higher frequencies, the GUVs exhibit mostly passive Brownian motion (Supplementary Movie 3). In regimes where $\rm f > 10~kHz $, the electric double layer at the particle surface does not have the time to build up, and the imposed AC electric field time scale $\rm \tau_f = 1/f$ is much faster with $\rm \tau_f \ll \tau_c$, where $\rm \tau_c = \frac{\lambda~a}{D_{ions}}$, with $\lambda$, a and $\rm D_{ions}$ being the Debye length, particle radius and the diffusion coefficient of the ions in solution, respectively \cite{Shilov2000,Squires2004}. In our system, for a frequency of 40 kHz and a particle radius average of 4.5 $\rm \mu m$, $\rm \tau_f = 25~\mu s$, with $\rm \tau_c = 4~ ms $, thus the periods of the AC field are much faster than the double layer can follow (Supplementary Note 1, Supplementary Video 3). Additionally, we observed deformation \cite{Dimova2007} of the vesicles all over the phase diagram except in the regions where the vesicles break (Supplementary Fig.S4, Video 4, and Note 1). 

%The proximity of the electrodes to the particles plays a crucial role in their motion. The timescale at which the diffuse cloud of counterions builds up near the electrode is $\rm \tau_{c,E} = \frac{\lambda ~ L}{D_{ions}}$, with L being the cell thickness. Similarly to the particle's double layer behaviour, when $\rm \tau_{f} \ll \tau_{c,E}$ the polarization of the double layer of the electrode does not follow the AC periods, with $\rm \tau_{c,E} = 4~ms$, with no effect on the particle motion \cite{Bazant2004}. 

Interestingly, when applying intermediate frequencies below 30 kHz down to 5 kHz, we observed active motion at amplitudes between 3-10 V$_{pp}$ (Supplementary Video 5). At this frequency range, surface charge polarization takes place via induced charge electroosmosis (ICEO) \cite{Bazant2004,Squires2004}. As we will show later in more detail, the active motion of the vesicles is highly dependent on the phase separation of the lipid domains of the vesicles and on their surface properties \cite{Ristenpart2007, Ma2015, Ma2015a}.
We perform control experiments with non-phase separated vesicles (i.e., only one type of lipid) which do not exhibit active motion due to the lack of asymmetry on the vesicle membrane (Supplementary Movie 6). We also perform control experiments at low salt concentrations $\rm (10^{-6} M)$, causing the consequent decrease of the EHD flows magnitude \cite{Yang2019}. Finally, we decrease the amount of charged dye, which leads to a decrease in the EHD magnitude. In all of the aforementioned cases, the motility of the Janus GUVs was cancelled or considerably reduced (Supplementary Note 2, Supplementary Fig.S5, Fig.S6).

\subsection*{Active motion via ICEO of Janus vesicles}

\begin{figure*}[h]
\includegraphics[width=\linewidth]{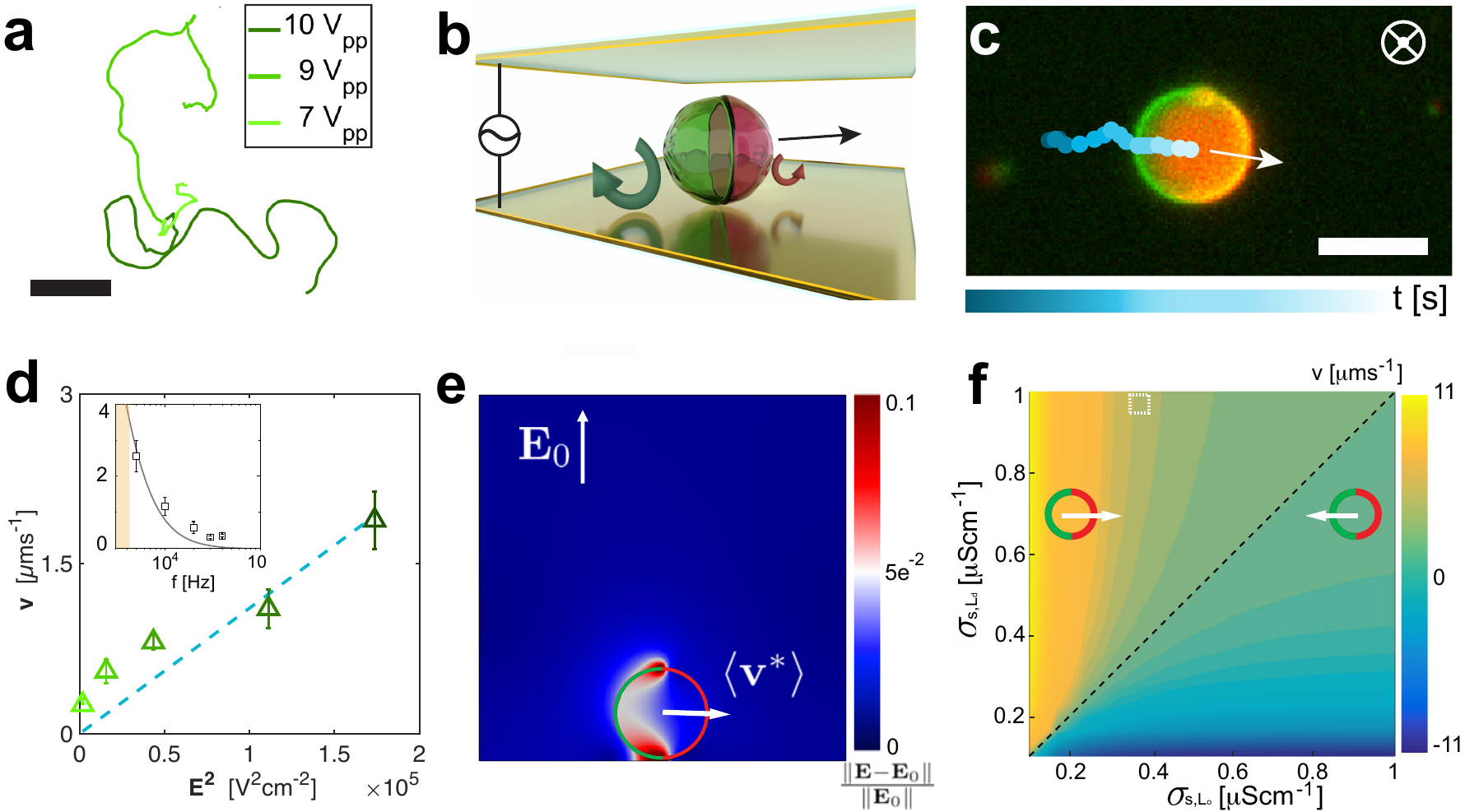}
\caption{\textbf{Dynamical characterization of Janus GUVs under AC fields}. ~\textbf{a,} Typical trajectories of active GUVs at increasing field strength of 7, 9 and 10 Vpp at 10 kHz.~\textbf{b,} Scheme of the Janus GUVs in the experimental cell exhibiting the EHD flows at each hemisphere (red and green arrows) confirmed experimentally by tracers. The black arrow depicts the direction of the motion.~\textbf{c,} Fluorescence image of an active GUV with its trajectory overlayed at 10 kHz and 9Vpp. The white arrow indicates the direction of motion, perpendicular to the applied field $\rm v \propto E^2$. The scale bar indicates 10 $\rm \mu m$.~\textbf{d,} Janus GUV velocity $\rm v$ as a function of the square of the applied field. Open symbols represent experimental data. The error bars indicate the standard deviation of more than 30 particles. The solid line is a linear fitting. The inset depicts the velocity as a function of frequency at 9 Vpp. The dashed line is a guide to the eye. \textbf{e,} Simulated colour plot obtained from COMSOL of the relative electric field amplitude normalised to the applied field $\mathbf{E}_0$. The white arrow is the average velocity vector $\langle \mathbf{v}\rangle$. The vesicle diameter is $\rm 10 \mu m$, the separation from the electrode $\rm 100 nm$, and the thickness of the membrane is 5 nm.~\textbf{f,} Simulation of the horizontal component of the average velocity $\rm v$ as a function of $\rm \sigma_{s,L_d}$ and $\rm \sigma_{s,L_o}$, including a prefactor $\beta=$0.02 to match our experimental data. The black dashed line is the value for which $\langle v_x\rangle$ = 0, and the white open symbol represents the velocities of our experiments.}
\label{fig:fig3}
\end{figure*}

We now focus on the active motion of Janus GUVs and rationalize the propulsion mechanism of the vesicles. In Fig.\ref{fig:fig3}a, we depict the typical trajectories of an active Janus vesicle with increasing voltage ($\rm V_{pp}$) at a fixed frequency of 10 kHz. Geometrically or compositionally asymmetric hard colloids  exhibit active motion via induced charge electroosmosis (ICEO) under AC electric fields in the kHz range between parallel electrodes \cite{Ganwal2008, Ma2015}. The oscillating electric field \textbf{E} applied in the vertical direction produces recirculating electrohydrodynamic (EHD) flows around the particle due to the polarization of the surface charges of the vesicle and its interaction with the bottom electrode \cite{Ristenpart2007}. For a Janus particle with two dielectric hemispheres, these flows are unbalanced at each hemisphere, causing net motion of the particles in the plane parallel to the electrode \cite {Buttinoni2022}. Importantly, the magnitude and direction of the EHD flows depend on the contrast between the surface conductivity of the particle ($\sigma_s$) and the conductivity of the media ($\rm \sigma_h$) \cite{Yang2019}. Similarly, the Janus GUVs exhibit active motion via EHD flows, swimming with the red hemisphere in front as shown in Fig.\ref{fig:fig3}b-c, with each lipid phase acting as a dielectric hemisphere with contrast in surface dielectric properties \cite{Buttinoni2022}. 

To rationalize this behaviour, we confirm the presence of the EHD flows around vesicles experimentally by using small tracer particles ($\rm \approx 1 \mu m$ diameter) to visualize these flows (Supplementary Fig.S7a,  Supplementary Video 7). Both EHD flows are repulsive (from the top to the bottom of the particle and moving away), and their magnitude decreases with the distance from the particle. The green phase containing DPPC+NBDPE exhibits EHD flows with a slightly greater magnitude than the red phase containing DOPC+NBDPE+RhPE, which is consistent with the direction of the motion. The difference in surface charge and conductivity among both phases of the Janus GUVs is the result of the presence of negatively charged fluorescent lipids in the two phases. We measure $\rm \zeta_{L_o} = -33 mV \pm 5$, and $\rm \zeta_{L_d} = -45 mV \pm 7$, measured by DLS for reference GUVs formed by each one of the main lipids (see more details in Methods). NBDPE dye dissolves in both membrane domains, while RhPE only partitions into the $\rm L_{d}$ phase, leading to a higher negatively charged surface in the $\rm L_d$ domains due to the presence of both charged dyes. The increase in surface charge magnitude leads to an increase of the zeta potential, thus increasing the local magnitude of surface conductivity ($\rm \sigma_{s,L_d} = 1~\mu S cm^{-1}$, $\rm \sigma_{s,L_o} = 0.35~\mu S cm^{-1}$) and decreasing the magnitude of the EHD flows \cite{Yang2019}. This contribution from the $\rm \sigma_s$ leads to a decrease in the EHD flow velocity around the vesicle in the regions where the $\rm L_d$ (red) phase is present, with a dominant flow coming from the $\rm L_o$ (green) hemisphere.  

In Fig.\ref{fig:fig3}d, we show the velocity $\rm v$ of the GUVs as a function of the applied field calculated from the average of the instantaneous velocities of each trajectory at each condition. The velocity exhibits a quadratic dependence of $\rm v$ with the electric field E as $\rm v \propto E^2$, consistent with the ICEO propulsion mechanism \cite{Ma2014,Ma2015}. Moreover, the velocity decreases with increasing frequency as $\rm v \propto~ f^{-1}$, as a result of the slower double-layer polarization at the particle surface with polarization time $\rm \tau_{c}$, with respect to the field frequency applied. This is consistent with the absence of motion at higher frequencies previously characterized in the phase diagram. The Janus GUV velocity has a distinct dependence on the phase separation asymmetry. In particular, we found that Janus vesicles with a higher percentage of $\rm L_d$ (green) phase will be faster, with velocities reaching thermal noise when the asymmetry decreases (Supplementary Fig.S7b). 

To gain further insight into the behaviour and motion of a bilipid Janus vesicle under an AC electric field, we run 2D electrostatic simulations using the finite element method-based commercial software COMSOL Multiphysics. We used an optimized model from the standard used for Janus hard colloids. We consider a thin membrane (5 nm) of 5 $\rm \mu m$ radius, with two hemispheres corresponding to each phase domain, and dispersion media with the same dielectric permittivity $\rm \epsilon_h$ and conductivity $\rm \sigma_h$ inside and outside the membrane. The dielectric constants of the two membranes are set to $\rm \epsilon_s=$10 \cite{waver2003}, and the conductivities are set to $\rm \sigma_{s,L_d} = 1~\mu S cm^{-1}$, $\rm \sigma_{s,L_o} = 0.35~\mu S cm^{-1}$. Under a constant background electric field, $\mathbf{E}_0$, the total electric field $\mathbf{E}$ is solved in Fig.\ref{fig:fig3}e is a colour plot representing the normalized relative field $\lVert\mathbf{E}-\mathbf{E}_0\rVert/\lVert \mathbf{E}_0\rVert$. 
Next, using the total electric field, the average velocity can be computed, which we define as $\langle \mathbf{v}^*\rangle = \frac{3\varepsilon_0\varepsilon_hV_p \langle\nabla\lVert\mathbf{E}\rVert^2\rangle}{6\pi\eta R}$, where $\rm V_p$ is the volume of the particle, $\varepsilon_0$ is the free-space electric permittivity, $\eta$ is the viscosity of the host medium and $R$ is the radius of the particle. Here $\langle .\rangle$ denotes volume averaging and is carried out over the entire simulation domain. The relation between the average swimming speed $\langle \mathbf{v}\rangle$ and the seed velocity is $\langle\mathbf{v}\rangle = \beta\langle\mathbf{v}^*\rangle$, where $\beta=\beta'\rm K(\omega)$ is the Clausius-Mosotti factor for a thin membrane solution ($\rm K(\omega)\approx 0.9$) and $\beta'$ is a prefactor that accounts for effects inherent to the experimental setup that are not considered in the simulation, such as sources of loss (see details in Supplementary Note 3).

Note that with our definitions, the dielectrophoretic force is $\mathbf{F}_\mathrm{DEP} = \beta\langle\mathbf{v}^*\rangle$, the drag force is $\mathbf{F}_\mathrm{drag}=-6\pi\eta R\langle\mathbf{v}\rangle$. For a particle in a viscous medium travelling at a constant speed, the dielectrophoretic force is opposed to the drag force: $\mathbf{F}_\mathrm{DEP} = -\mathbf{F}_\mathrm{drag}$ and the average swimming speed $\langle\mathbf{v}\rangle$ can thus be retrieved consistently with our formula. From the simulation we observe that the horizontal component is the dominant on the average seed velocity, such that $v=\lVert\langle\mathbf{v}\rangle\rVert\sim\langle\mathbf{v}.\hat{x}\rangle$, where $\hat{x}$ is a unit vector in the direction of $\mathbf{v}$ \cite{Jones1995}. Using the experimental conditions in the simulation, we obtain a total velocity of $v= 1.5~\mu m$ with $\beta=0.2$, whose magnitude and direction agree with the experimental data at 10 kHz and 8 $\rm V_{pp}$. Moreover, the total velocity of the particle is very sensitive to small variations in contrast to the surface conductivity of each hemisphere $\rm \sigma_{s,L_o}$ and $\rm \sigma_{s,L_d}$. We evaluate the magnitude and direction of Janus GUV velocities varying the $\rm \sigma_s$ of both hemispheres (Fig.\ref{fig:fig3}f), with no effective swimming velocity when there is no conductivity contrast amongst hemispheres (Supplementary Fig.S8). This parametric representation offers promising guidelines for designing Janus GUVs with a tailored velocity modulation depending on their charge surface properties and asymmetry.

\subsection*{Phase separation dependent dynamics}

\begin{figure*}[h]
\includegraphics[width=\linewidth]{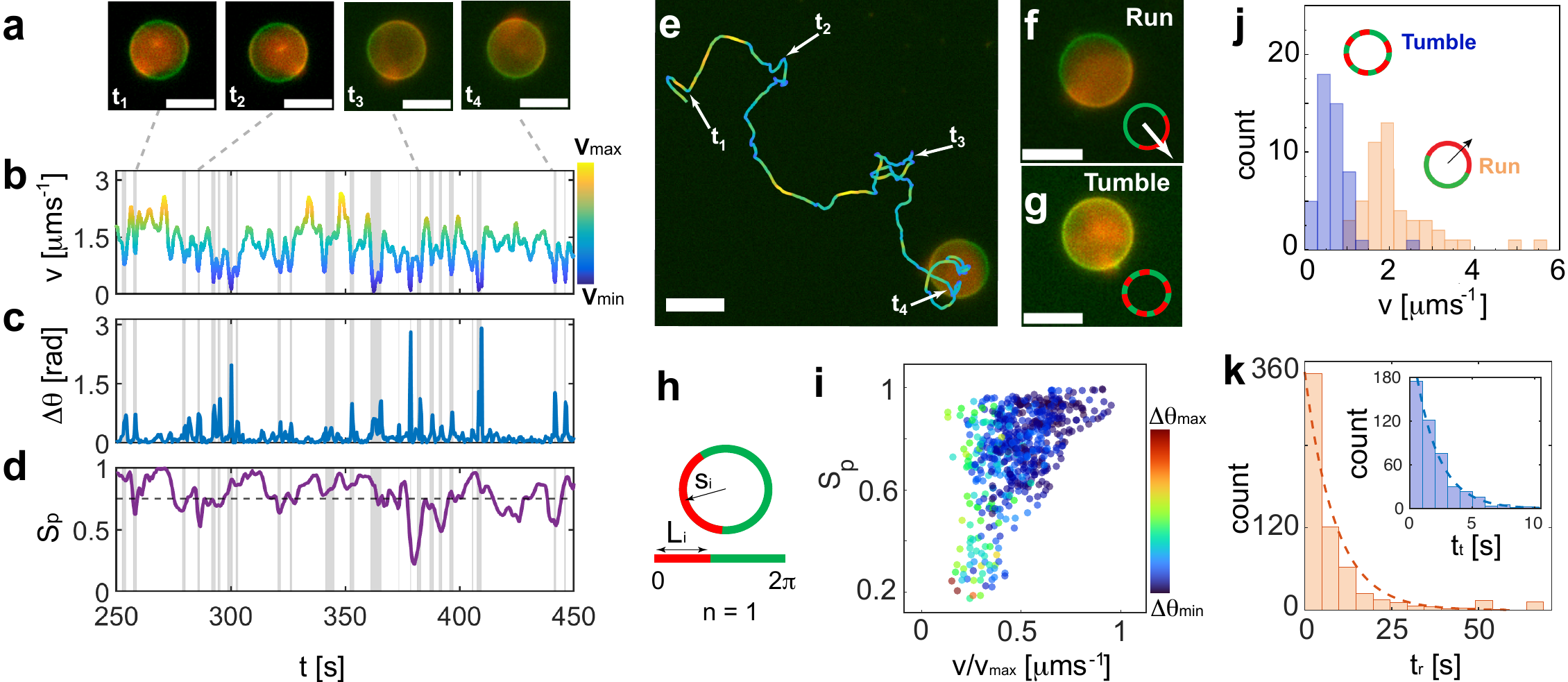}
\caption{\textbf{Run and tumble-like motion of Janus GUVs}~\textbf{a,} Fluorescence pictures of example tumble events ($\rm t_i$) due to membrane domain mixture, and loss of asymmetry. Scale bars represent 15 $\rm \mu m$. Evolution as a function of time of the ~\textbf{b} instantaneous velocity $\rm v$ as the displacements between two frames with dt = 0.4 s, ~\textbf{c} change in orientation $\rm \Delta \theta$  between two frames with dt = 0.4 s, and ~\textbf{d} order parameters $\rm S_p$ for the GUV in panels ~\textbf{a,f}. The dashed line in ~\textbf{d} represents the threshold below which the system is measured as disordered $\rm S_p<$0.75. The vertical grey areas indicate the tumble events detected by overlapping a local minimum in $\rm v$, a maximum in $\rm \Delta \theta$, and minima in $\rm S_p$.~\textbf{e,} Snapshot of a fluorescent microscopy video of an Active Janus GUV, overlaid with its trajectory at 10 kHz and 9 $\rm V_{pp}.$ The tumble events are indicated as $\rm t_i$. The colour code indicates the velocity evolution. Screenshots of~\textbf{f,} run and~\textbf{g,} tumble events of a Janus GUV. The inset depicts a schematic representation of the phases arrangement. Scale bars are $\rm 15 \mu m$.~\textbf{h,} Scheme of a 2D projection of a fully phase-separated GUV (left) and the parameters identified for calculating the order parameters $\rm S_p$ being $\rm s_i$ the orientation of the vector and $\rm L_i$ the length of the domain segment.~\textbf{i,} Dependence of $\rm S_p$ with normalized velocity. The colour code indicates the associated $\rm \Delta \theta$~\textbf{j,} Histogram of the mean velocities of the runs (orange) and tumbles (blue) after analysis of 30 trajectories.~\textbf{k,} Distributions $\rm \psi$ of tumble-time $\rm t_t$ and run-time $\rm t_r$ drawn from exponential fits as $\rm \psi_r \propto 1/\overline{t}_re^{-t_r/\overline{t}_r}$ and $\rm \psi_t \propto 1/\overline{t}_te^{-t_t/\overline{t}_t}$ respectively, where $\rm \overline{t}_r$ and $\overline{t}_t$ are the mean run-time and mean tumble-time, respectively.}
\label{fig:fig4}
\end{figure*}

Inspecting the motion of the vesicles more closely, we find unexpected dynamical behaviour that we relate to the intrinsically reconfigurable phase-separated hemispheres of the GUV. Interestingly, the fluid nature of the bi-lipid membrane allows for the periodic mixing of both lipid phases under the effect of the AC electric field, leading to transient dynamics states that we call in this work \textit{run} (phase separated GUVs) and \textit{tumble} (loss of asymmetry) events, due to their similarity to the characteristic navigation strategy performed by bacteria \cite{Seyrich2018, Thiel2012} (Fig.\ref{fig:fig4}, and Supplementary Movie 8). This inherent property is in stark contrast with the discrete geometry exhibited by hard Janus particles and other colloidal assemblies, where the defined material boundaries impede any geometrical, and thus, dynamical reconfiguration. While in the absence of the electric field, both phases are thermodynamically separated, the AC electric field coupled with the interaction with the substrate strongly affects the phase separation, generating punctual mixing  events \cite{Dimova2009}.

We characterize the presence of run and tumble events at 5 - 10 kHz and 7 - 9 Vpp, with optimized tumble identification algorithms commonly used for the quantification of bacterial dynamics \cite{Seyrich2018}. In particular, for each trajectory, we monitored the evolution of the instantaneous velocity $\rm v$, and the orientation variation of $\rm \Delta \theta$ as a function of time (Fig.\ref{fig:fig4}a,b,c). We locate local velocity minima and flag a potential tumble if the velocity decreases by more than 70\% relative to the value at the corresponding local minimum with respect to neighbor points. We also evaluate the angular displacement, $\Delta \theta$, and compare it to a threshold obtained from the distribution of angular displacements. We confirm a tumble event only when both conditions are met at the same time point (see Supplementary Note 4). This drop in velocity is related to the phase domains of the vesicle mixing, which leads to asymmetry loss and the loss of activity, meanwhile, there is an increase in the jittering of the particle in place during the tumbling events (Fig.\ref{fig:fig4}e,f,g). 
To have a better insight into the membrane phase domain structure during these processes, we quantify an order parameter $\rm S_p$ (Fig.\ref{fig:fig4}d) that measures the spatial distribution of red-labelled regions along the vesicle's perimeter. We integrate both the angular position $\rm s_i$ and the length of each segment L$\rm _i$ for the red domains, serving effectively as an order parameter for the alignment of labelled regions (Fig.\ref{fig:fig4}h). This is done by detecting red pixels on the perimeter of the 2D projection from the fluorescence images (Supplementary Note 4). Each segment's spatial distribution is translated into a vector in the 2D plane. The direction of this vector is aligned with the segment's angular position, and its magnitude is determined by the segment's length \( \rm L_i \). Formally:
\begin{equation}
\mathbf{s}_i = L_i \begin{bmatrix} \cos(\theta_i) \\ \sin(\theta_i) \end{bmatrix}
\end{equation}
The cumulative contribution of all individual segment vectors is obtained by computing their weighted average, resulting in a weighted-polar-vector $\mathbf{S}_p$,
\begin{equation}
\mathbf{S}_p = \frac{\sum_{i} \mathbf{s}_i}{\sum_{i} L_i},
\end{equation}
the magnitude of the $\rm S_p = \vert \mathbf{S}_p \vert$ provides the overall alignment strength of the red-label signal and therefore delineates the dominant orientation of the red-labelled regions on the vesicle. If  $\rm S_p > 0.75 $, the Janus vesicle asymmetry is related to a transition to an ordered state, while if $\rm S_p<0.75$ the asymmetry is lost. The tumble events coincide with an overlap in a drop in $\rm v$, the increase in $\rm \Delta \theta$, and an order parameter $\rm S_p<0.75$. We found a correlation of the $\rm S_p$ with $\rm v$ and the $\Delta \theta$ (Fig.\ref{fig:fig4}i, Supplementary Fig.S9) where the highest values of $\rm S_p $ coincide mostly with high values of $\rm v$ and low $\Delta \theta$, while the tumbling events show a distinct decrease of $\rm S_p $ and $\rm v$ and an increase of $\Delta \theta$. It is important to note that since the analysis is performed on a 2D projection of a 3D image, we have inherent noise due to the underestimation of certain phase domain configurations. Through this methodology, we analyze the spatial distribution and patterning of labels on vesicles, providing insights into their spatial arrangement and dynamical behaviour (Supplementary Fig.S10 and Fig.S11 for more examples). 

Looking at the typical velocities of run and tumble after the classification of each event, we find that the runs are distinctly faster (Fig.\ref{fig:fig4}j). We observe run and tumble times ($\rm t_t$ and $\rm t_r$, respectively) distributions following an exponential decay of tumble-time $\rm t_t$ and run-time $\rm t_r$ given by the exponential fit as $\rm \psi_r \propto 1/\overline{t}_re^{-t_r/\overline{t}_r}$ and $\rm \psi_t \propto 1/\overline{t}_te^{-t_t/\overline{t}_t}$ respectively, where $\rm \overline{t}_r$ is the mean run-time and $\overline{t}_t$ is the mean tumble-time (Fig.\ref{fig:fig4}k), as expected for run and tumble motion \cite{Karani2019} with both quantities being randomly distributed in the trajectory, a reminiscent feature of bacterial swimming patterns \cite{Min2009}.

\begin{figure*}[h]
\includegraphics[width=\linewidth]{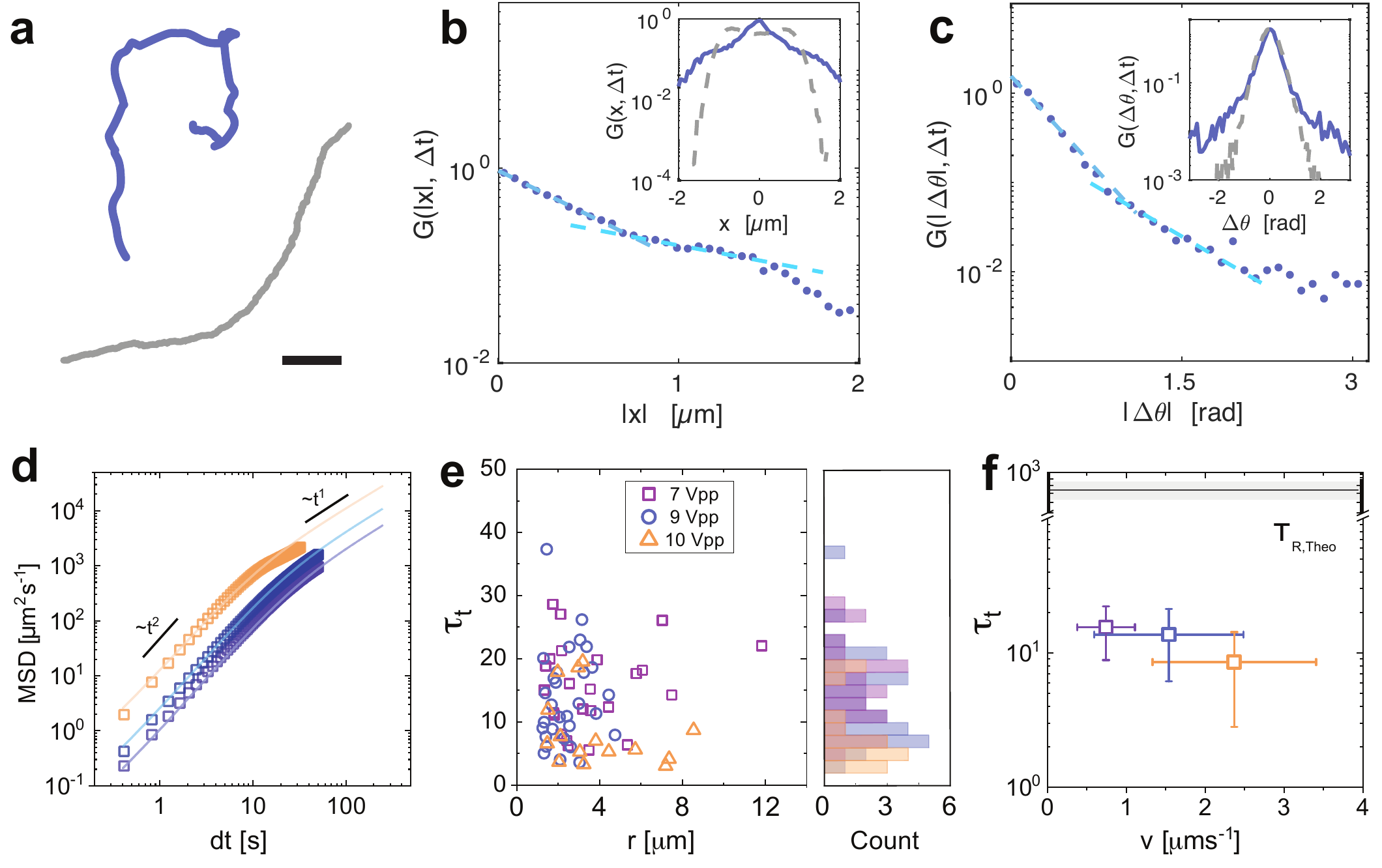}
\caption{\textbf{Statistical analysis and field dependent reorientation}.~\textbf{a,} Comparison of the experimental trajectory (blue) at 10 kHz and 9Vpp with $\rm \tau_{R} = 11 \pm 4~s$ and simulated trajectory with ABP model (grey) of a particle with analogous overall $\rm v = 2~\mu ms^{-1}$  and $\rm \tau_{R,Theo} = 180~s$. The scale bar depicts 20 $\rm \mu m$. The probability distribution function of ~\textbf{b,} the displacements x and~\textbf{c,} the orientation $\rm \Delta\theta$. The blue solid line represents the experimental data for 30 particles at 10 kHz and 9 Vpp and the light blue dashed lines represent the exponential fit of $\rm G \propto 1/e^{x}$ at small and large displacements or angles respectively, where x represents the data plotted on the x-axis. The grey dashed line in the inset  represents the simulated data
~\textbf{d,} Mean square displacements of particles at increasing voltage conditions and fixed frequency (10 kHz). The open symbols represent the ensemble average for each condition, and the solid lines represent the fitting for long times.~\textbf{e,} Reorientation time $\rm \tau_{t}$ obtained from the individual mean-square displacements, as a function of GUV radius r. The right plot indicates the histogram of the data points.~\textbf{f,} Decreasing reorientation time $\rm \tau_{t}$ with increasing velocity (thus, field applied) for the particles in \textbf{d}. The open symbols represent the experimental data. The solid line and the shaded square represent the theoretical prediction, $\rm \tau_{R} = D_{R}^{-1}$ using Einstein-Stokes $\rm D_{R}=\frac{kT}{8 \pi \eta r^3}$. }
\label{fig:fig5}
\end{figure*}

We perform statistical analysis for both spatial $\rm G(|x|,t)$ and orientational $\rm G(|\Delta \theta|,t)$ displacements at short time scales (dt = 0.4s), and compare these results using Active Brownian particle (ABP) simulations \cite{Volpe2014} for particles with the same characteristic as the Janus GUVs ($\rm v$, diameter, and rotational diffusion $\rm D_R$ associated to the particle size) as shown in Fig.\ref{fig:fig5}a,b,c (Supplementary Note 5). For the experimental data, we observe a non-Gaussian behaviour with a transition from a first exponential phase at small $\rm |x|$ and $|\Delta \theta|$, followed by an exponential tail at longer displacements. Both exponential phases are fitted with the expression $\rm G \propto 1/e^{x}$ where x represents the data plotted on the x-axis. The exponential tails in $\rm G$ are characteristic of dynamical heterogeneity such as in glassy phases or bacterial dynamics \cite{Chaudhuri2007, Seyrich2018}, where faster and slower events coexist. This is the case for our system in constant transition between runs and tumbles, where $\rm \Delta \theta_t \gg \rm \Delta \theta_r $, and $\rm v_t \ll v_r$. Interestingly, we recover also exponential behaviour at short displacements, as opposed to the expected Gaussianity, hinting at heterogeneous dynamics at short times, dominated by the large spatial and angular fluctuations when both phases of the GUV are not fully separated when overcoming a tumble. There is also a distinct deviation from the ABP model behaviour, where for $\rm G(x,t)$ we expect a broader and flatter distribution with a double peak in the ballistic regime \cite{Zheng2013}, and for $\rm G(\Delta \theta,t)$ a single Laplace distribution, where $\rm G(\Delta \theta,t)\propto e^{\frac{-|x|^b}{c}}$ \cite{Lemaitre2023}. Overall, the presence of the run and tumble events combined with the constant fluctuations of both phases of the GUVs lead to a non-constant swimming force, reflected in the velocity variations and the heterogeneity observed at short displacements. 

Furthermore, we measure the mean square displacements (MSD) with trajectories of at least t = 150 s for increasing voltage conditions and fixed frequency at 10 kHz (Fig.\ref{fig:fig5}d). We extract reorientation times by fitting the expression of the MSD at long times for ABP \cite{Mestre2020,Bailey2022}, obtained from the relaxation from ballistic to diffusive in the MSD. While the expected thermal reorientation time $\rm \tau_R$ for a spherical particle of 10 $\mu m$ diameter is $\rm \approx 10^3$ s, where $\rm \tau_R = D^{-1}_R = \varsigma / kT$,  ( $\varsigma$ being the drag coefficient), we find a drastic difference given by the tumbling-driven reorientation, where $\rm \tau_R \approx 100 \tau_t$. Notably, the extracted $\rm \tau_t$ from the MSD does not exhibit an effect of the size (Fig.\ref{fig:fig5}e), which is in discrepancy with the Einstein-Stokes equation $D_R = k_B T/ \varsigma $, supporting that $\rm \tau_t$ is decoupled from $\tau_R$, and it is associated with an enhanced reorientation due to the tumble events or $\rm \tau_t$. In addition, $\rm \tau_t$ depends on the activity decreasing as $\rm v$ increases, hinting at a strong correlation between the tumbling events with the applied field (Fig.\ref{fig:fig5}f). The consistent data deviation is given by the broad asymmetry population of the phase-separated vesicles, which has an important influence on their instantaneous average velocities, with faster vesicles when the $\rm L_d/L_o < 1$ as previously mentioned (Supplementary Fig.S7).

Overall, the constant but random transition between running and tumbling events of the GUVs arises from the interaction of the membrane with the rough topography of the bottom electrode. Topographic roughness arises from ITO deposition which generates point defects perturbing punctually the electric field leading to the perturbation of the phase separation, as previously shown by Dimova et al. for passive GUVs in co-planar electrodes \cite{Dimova2009}.

\subsection*{Motion and functionality design}
\begin{figure*}[h]
\includegraphics[width=0.7\linewidth]{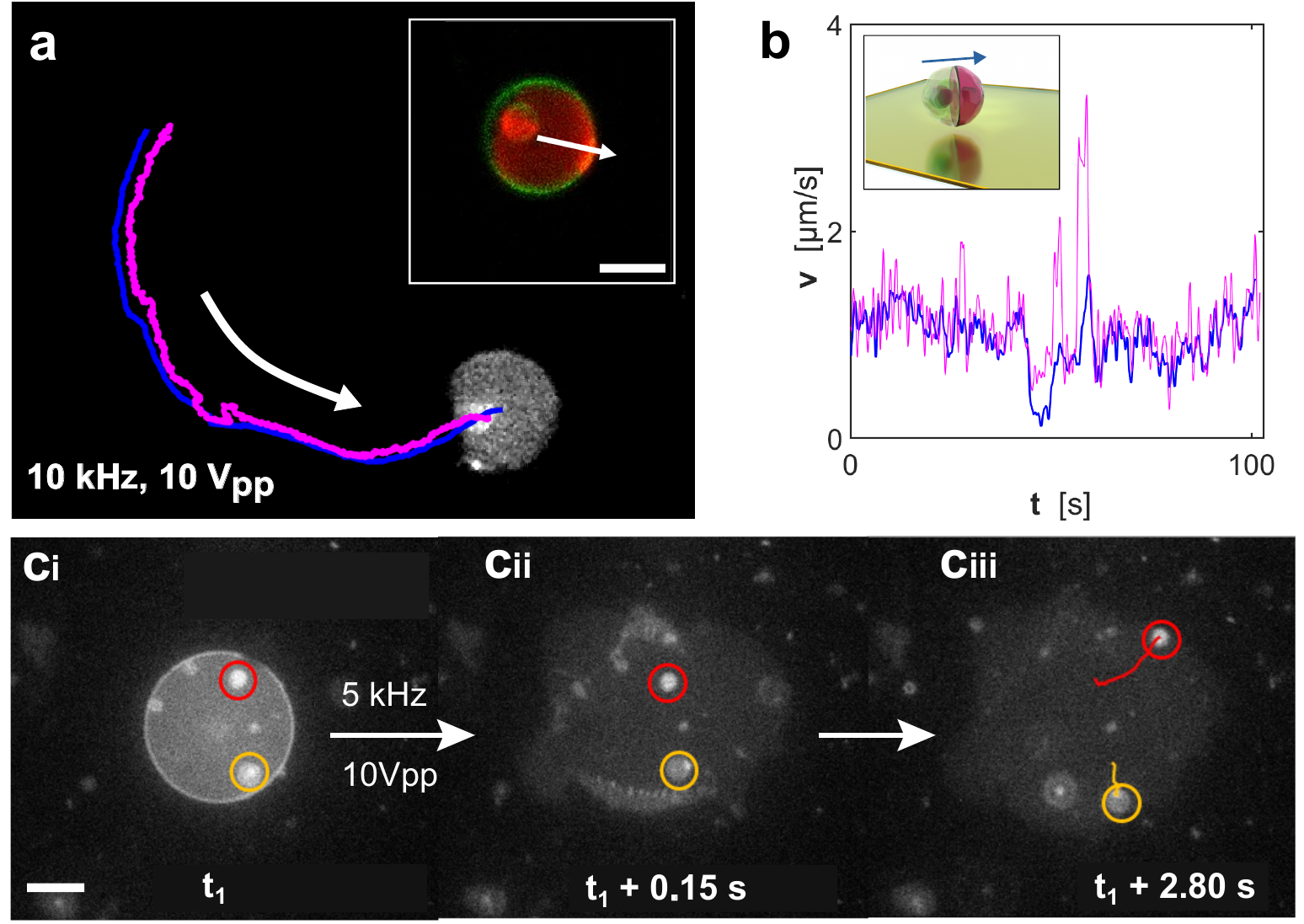}
\caption{\textbf{a,} Fluorescence image in BW representing the $\rm L_d$ phase of a GUV, containing a smaller GUV. The pink and blue lines represent the trajectory of the small inner and big outer GUVs, respectively. The inset represents the confocal image of the green and red channels. The scale bar depicts 10 $\rm \mu m$. \textbf{b} Velocity as a function of time for the inner and outer GUVs. The inset represents a schematic representation of the system. $\rm \bf{c_i}$, $\rm \bf{c_{ii}}$, $\rm \bf{c_{iii}}$ Time sequence of on-demand cargo release by bursting the vesicles upon decreasing the frequency to 5 kHz. The scale bar depicts 10 $\rm \mu m$.}
\label{fig:fig6}
\end{figure*}

In contrast to common hard colloidal active particles, the compartmentalised nature of GUVs allows versatile loading of cargo inside the compartment \cite{Baastiaan2017}. Here we demonstrate that encapsulation of cargo and further on-demand delivery is possible by utilizing the response of the active GUVs to the applied AC electric field.
The GUV production process with electroformation, can be adapted to end up with bigger populations of GUVs containing smaller vesicles. We took advantage of this to demonstrate a great feature of the vesicles: their utility as a cargo transport and on-demand release system. For this purpose, we use the same configuration for active motion as described in the previous sections with parallel electrodes. Upon application of the AC-field (10 kHz, 10 $V_{pp}$) the larger vesicle ($\rm r = 6 \pm 3.6 \mu m$) containing the smaller vesicle ($\rm r < 2 \mu m$) inside as cargo exhibits active motion (Fig. \ref{fig:fig6}a-b, Supplementary Movie 9). We confirmed that the motion of the large surrounding vesicle was not impaired by the inner cargo, and the latter exhibits Brownian noise inside the container. During transport, we impose controlled release of the cargo by applying lower frequencies (1-5 kHz) as previously described, to burst the bilipid membrane and release the cargo on-demand, as shown in figure Fig. \ref{fig:fig6} (Supplementary Movie 10).
Ultimately, we explored the dynamics of the variety of architectures resulting from the electroformation (Fig.\ref{fig:fig1}, Supplementary Fig.S12, Supplementary Movie 11). In particular, we found intrinsic chiral dynamics of 'patchy Janus GUVs' in which phase separation is not completed. In this case, we consistently found non-completed $\rm L_o$, in which two $\rm L_o$ patches (green) of different sizes were on the membrane.  The resulting architecture favours chiral trajectories since the forces experienced on the hydrodynamic centre of the GUV are not centred leading to a torque \cite{Wang2020}. In addition, 'dumbbells' formed by two lobes containing each a distinct lipid phase, exhibit active behaviour characteristic of this type of geometries \cite{Ma2015}. Contrary to the Janus GUVs, the two separated phases of the dumbbell do not mix, and the dynamics that they exhibit are expected for dumbells of such size. Their properties as soft containers for programmable release coupled with their phase-separation-dependent motion demonstrate the potential of these cell-mimetic units as motile and functional synthetic membranes.

\section*{Discussion}

In this study, we elucidate the rich dynamics of phase-separated Janus GUVs which serve as compartmentalized artificial microswimmers with inherent reconfigurable dynamics. We found transitions between run-and-tumble states, driven by the fluidity of the bilipid layer. In particular, we demonstrate that Janus GUVs are motile under AC electric fields applied between parallel electrodes, a phenomenon explained through Induced Charge Electroosmosis (ICEO) and supported by numerical calculations. While the activity is given by external actuation, the reconfiguration is purely intrinsic to the ability of the phase domains of the lipid membrane to mix via constant interaction with the surface while the electric field is applied. We quantify the run and tumbles and associate these distinct dynamics states to the domain structure of the lipids vesicle, undergoing distinct transitions from full phase separation to complete mixing. By quantifying their dynamics, we demonstrate an enhancement of the reorientating time attributed to the tumbling events. The variation of each phase domain, and therefore asymmetry of the Janus GUV, leads to changes in swimming speeds, revealing a key parameter to design active lipid vesicles with diverse motility. Finally, we demonstrate the potential for coupling these reconfigurable dynamics with transport-release functionalities.

The use of active GUVs with phase-separation-dependent motion presents a step forward towards engineering cell-mimetic microswimmers. The design presented in this work depicts a purely Markovian telegraph process \cite{Seyrich2018}, with transitions between two states (runs and tumbles). The temperature-driven phase separation of the lipid hemisphere unlocks an unprecedented tool for the design of reconfigurable and internally driven temperature-tactic dynamics. This feature allows for the design of non-Markovian dynamics, where the interaction with the environment affects the motion of the vesicle, reminiscent of biological microorganisms \cite{Angelani2019}, aiming for membrane structure-dependent motility. The support of machine learning tracking algorithms will improve the recognition of tumble events of 2D images and therefore increase the understanding of the effect of the membrane domain conformation on the run and tumble events. In addition, their ability to serve as cargo-transport units will open up vast possibilities and interactions with synthetic and biological systems \cite{Wittmann2023}. Nevertheless, a higher degree of control over the size and composition of these active Janus GUVs and their surroundings is needed to investigate these questions. 

Future developments of this strategy are closely connected to progress in advanced artificial cells, featuring motion via chemical signalling as explored in membraneless compartments such as droplets \cite{Testa2021,Meredith2020}, navigation strategies in complex environments \cite{Khatami2016, Dou2019}, or collective behaviour with soft potentials \cite{Hopkins2023}. This brings us a step closer to mimicking the behaviour of highly evolved cellular microorganisms with engineered systems \cite{Boudet2021, Sjoerd2018, Perez2023}.

In conclusion, our experimental findings present phase-separation-dependent dynamics in Janus GUVs, exhibiting run-and-tumble behaviour due to inherent bilipid membrane properties. We identify the role of domain structures in shaping these transient states and characterize their non-size-dependent reorientation times. These results open avenues for programming motility and signalling based on membrane responses to external stimuli, establishing a benchmark for more intricate cell-mimetic microdevice designs.

\section*{Methods}
\subsection*{Materials}
Lipids: 1,2-dioleoyl-sn-glycerol-3-phosphocholine (DOPC), 1,2-dipalmitoyl-sn-glycerol-3-phosphocholine (DPPC), 1,2-dioleoyl-sn-glycerol-3-phosphoethanolamine-N-(lissamine rhodamine B sulfonyl) (RhPE) and 1,2-dipalmitoyl-sn-glycerol-3-phosphoethanolamine-N-(7-nitro-2-1,3-benzoxadiazol-4-yl) (NBDPE) were bought from Avanti Polar Lipids (USA). Cholesterol (highly purified) was purchased from EMD Millipore Corp (USA). Chloroform (ReagentPlus, $\geq$99.8$\%$, contains 0.5-1$\%$ ethanol as a stabilizer), Sucrose (BioUltra,$\geq$99.5$\%$) and Poly-(sodium 4-styrene sulfonate) (P-S-sulfonate) (average Mw $\approx$ 70'000, powder) and spacers Grace Bio-Labs SecureSeal imaging spacer (8 wells, 9 mm diameter, 0.12 mm thickness) were purchased from Sigma-Aldrich. Copper tape (double-sided, 12 mm x 16.5 m x 2 mm) was purchased from Distrelec. ITO-slides from Corning were used for electroformation (50*75*1.1 mm, Rs = 4-10$\Omega$) and AC-field experiments (25*25*0.7 mm, Rs = 9-15$\Omega$ and 25*25*1.1 mm, Rs = 4-8$\Omega$). MiliQ water was used and filtered with 0.2$\mu$m CA-membrane filters (ClearLine). 

\subsection*{Phase-separated Janus vesicle formation via electroformation}
Lipids dissolved in chloroform to 1mM (35:35:30 DOPC/DPPC/cholesterol molar ratio + 0.1$\%$ RhPE + 1$\%$ NBDPE)  were drop casted onto an ITO-coated glass plate to form a thin film inside the wells of a PDMS spacer and covered with a second ITO plate and clamped together. The whole cell was placed in a vacuum desiccator for at least 1 hour, up to overnight, to remove the remaining chloroform. Then, the wells were filled with 25 mM sucrose in filtered MiliQ solution, the electrodes were connected to a function generator (Keysight 33500B Waveform Generator Trueform), and the cell was placed on a heating plate at 60 °C, applying an electric field adapted from previous protocols for GUV formation \cite{Spanke2022}. The solution containing the GUVs was collected and stored in Eppendorf tubes at room temperature (25°C) and protected from light for a maximum time of one week. 
For DLS measurements, non-phase separated GUVs were made via the same electroformation protocol at as similar as possible compositions of each of the individual phases. It has been measured that the liquid-disordered ($\rm L_d$) phase contains approximately 1/3 of the cholesterol and the liquid-ordered ($\rm L_o$) phase contains the remaining 2/3 \cite{Veatch2003}. Therefore, the vesicles corresponding to the $\rm L_d$ phase contain  a molar ratio of 90:10 DOPC/chol +1$\rm\%$ NBDPE + 0.2$\rm\%$ RhPE, while the vesicles representing the $\rm L_o$ phase are formed with a molar ratio of 80:20 DPPC/chol +1$\rm\%$ NBDPE. Additionally, two control samples were made without dyes with molar ratios of 90:10 DOPC/chol and 80:20 DPPC/chol respectively.

\subsection*{Sample and AC electrics field}

To build the experimental cell, we first clean the ITO-coated glass slides (0.7 mm thickness) by sonicating them for 5 min in ethanol, then 5 min in MiliQ water, and finally drying them with compressed air. Then, we placed a  spacer (120 $\mu$m) on the conductive side of one of the ITO slides and filled it with 10~mg/ml P-S-sulfonate (PSS), and left it for 10 min. Then the remaining PSS was removed and the excess was rinsed 5 times with MiliQ water. A volume of the solution (7.4 $\mu$l) containing the GUVs was placed into the spacer and covered with an unfunctionalized ITO slide. Electrodes were attached to the cell with copper tape and connected to a function generator (Keysight 33500B Waveform Generator Trueform) to apply a sinusoidal AC-field ranging between 1-10 V$\rm_{pp}$ and 1-10 kHz. 

\subsection*{Microscope observation and tracking}
We first characterized the behaviour under a wide variety of different voltage conditions using an inverted optical microscope (Olympus IX71) to observe the GUVs labelled with fluorophores with a 40x objective and an LED light source (cool LED pE excitation system). Images were taken with a Canon EOS 77D camera, with exposure times of 30 $\rm s$, at which the trajectories became visible, allowing for determining the presence of active motion. For further quantification of the active motion and phase separation of the GUVs at specifically chosen field conditions, we use a confocal microscope (Zeiss LSM 980 Airyscan with camera Zeiss Axiocam 705 mono) in fluorescence mode. Before recording, an AC field of 40 kHz and 1 V$\rm_{pp}$ was applied to the sample and the amplitude was increased to 10 V$\rm_{pp}$ to ensure equilibration and electrostatic attraction of the GUVs to the substrate (otherwise density-matched). Then the frequency was decreased to 10 kHz and videos were recorded at 2-3.5 fps for 1000-1200 frames using both 475 nm and 555 nm wavelength LED light sources for excitation times of 50 - 100 ms each. To study the passive Brownian motion of the GUVs videos were taken at 10 fps with exposure times of 50 ms in each channel. We quantified the population of the phase-separated GUVs,  with the sample on non-conductive thin ($\#$1) glass microscopy slides. A 488 nm and a 561 nm wavelength laser were used to image both phases. A z-step of 0.62 $\mu$m was used, but due to the mismatch of the refractive indices, a correction factor k=n$_{water}$/n$_{air}$ was necessary for 3d reconstruction of the confocal stacks. The dynamical quantification was performed using pre-optimized particle tracking algorithms (MATLAB and ImageJ, Fiji)

\subsection{Statistical analysis}
We perform statistical analysis for 48 Janus vesicles of 10 $\rm \pm 4.5 \mu m $ at a fixed frequency of $\rm 10 kHz$ and increasing field amplitude, where the optimal active swimming behaviour was observed. In order to do this we calculate the probability of spatial and orientational displacements, as \cite{Chaudhuri2007, Lemaitre2023}

\begin{equation}
\begin{aligned}
 G_s(x,t) = \langle \delta (x - \mid x_i(t) - x_i(0)\mid\rangle \\
 G_s(\Delta \theta,t) = \langle \delta (\theta - \mid \theta_i(t) - \theta_i(0)\mid\rangle \\
\end{aligned}
\label{eq:gs}
\end{equation}

For $\rm G(x,t)$ the displacements in x and y for each trajectory are cumulated to increase statistics.

\section*{Data availability statement}
The data that support the findings of this study are available from the corresponding authors upon reasonable request. Source data are provided in this paper.

\section*{Code availability statement}
The code used in this study is available from the corresponding authors upon reasonable request.

\section*{Acknowledgements}
V. W and L. A. are thankful to Prof. M. Angelova, and Prof. H. Kellay for fruitful discussions. L.A. and V.W. acknowledge IdEx Bordeaux (France) for financial support. L. A. and V.W. thank Dr. M. Martin and Dr. E. Ducrot for access to confocal microscopy, and Dr. J.P Chapel for help with DLS measurements. D.M.F. thanks the Comunidad de Madrid and the Complutense University of Madrid (Spain) through the Atraccion de Talento program 2022-T1/TIC-24007.

\section*{Author Contributions}
Author contributions are defined based on the CRediT (contributor roles taxonomy) and listed alphabetically. Conceptualization: L.A; Formal analysis: L.A., A.B, D.M.F, V.W.; Funding acquisition: L.A; Methodology: L.A., G.W, V.W; Project administration: L.A.; Resources: A.B, L.A, J.C.B, V.W.; Software: A.B, D.M.F, V.W, L.A; Supervision: L.A.; Visualization: L.A., V.W.; Writing original draft: A.B., D.M.F, J.C.B, L.A., V.W.; Writing review and editing: V.W., J.C.B, E.D. and L.A.

\section*{Competing Interests}
The authors declare no competing interests.

\section*{Additional Information}
Supplementary Information is available for this paper. Correspondence and requests for materials should be addressed to laura.alvarez-frances@u-bordeaux.fr

\bibliographystyle{naturemag}
\bibliography{mainactive}

\newpage

\newpage

\newpage

\newpage

\newpage

\newpage

\end{document}

\nolinenumbers

\title{{\sc Supplementary Information for:}
\\%
Phase-separation dependent active motion of Janus lipid vesicles}

\author{V. Willems}
\affiliation{CNRS, Univ. Bordeaux, CRPP, UMR5031, 33600 Pessac, France}
%\email{laura.alvarez-frances@mat.ethz.ch}
\author{A. Baron}%
\email{alexandre.baron@u-bordeaux.fr}
\affiliation{CNRS, Univ. Bordeaux, CRPP, UMR5031, 33600 Pessac, France}
\author{D. A. Matoz-Fernandez}%
\affiliation{Department of Theoretical Physics, Complutense University of Madrid, Madrid, 28040, Spain}
\author{G. Wolfisberg}
\affiliation{Laboratory of Soft Living Materials, Department of Materials, ETH Zurich, 8093 Zurich, Switzerland}%
\author{E. R. Dufresne}
\affiliation{Laboratory of Soft Living Materials, Department of Materials, ETH Zurich, 8093 Zurich, Switzerland}%
\author{J.C. Baret}
\author{L. Alvarez}
\email{laura.alvarez-frances@u-bordeaux.fr}
\affiliation{CNRS, Univ. Bordeaux, CRPP, UMR5031, 33600 Pessac, France}%

%\pacs{Valid PACS appear here}

%\begin{widetext}
\onecolumngrid
\parindent 0mm
\bigskip

\maketitle

\section*{Supplementary Movies}

\textbf{Supplementary Movie 1:} Phase-separated passive giant unilamellar vesicles (GUVs) exhibiting Brownian motion in the absence of an external AC field. RhPE segregates into the liquid-disordered phase (red) and NBDPE is distributed in the whole vesicle (green). Scale bars indicate 50 µm in example 1 and 20 µm in example 2.

\textbf{Supplementary Movie 2:} Example of a vesicle bursting in real time upon lowering the frequency of the applied AC field from 10 kHz to 1kHz. The scale bar indicates 25 µm.

\textbf{Supplementary Movie 3:} Vesicle behaviour at high frequencies (30 - 40 kHz) during sample equilibration. Vesicles do not exhibit active notion via ICEO at these frequencies. The scale bar represents 40 µm.

\textbf{Supplementary Movie 4:} Examples of vesicle deformation upon application, changing or removal of the AC field. The scale bars in all examples are 50 µm.

\textbf{Supplementary Movie 5:} Phase-separated vesicles exhibiting active motion via ICEO at AC field frequencies of 5 - 10 kHz. The scale bars in all examples indicate 50 µm.

\textbf{Supplementary Movie 6:} Example of a vesicle with one lipid type lacking asymmetry and this active motion under AC electric field of 10 kHz and 9 $\rm V_pp$. The scale bar is 15 µm.

\textbf{Supplementary Movie 7:} Small tracers being pushed by EHD flows around the vesicle, thus confirming the motion of the vesicles is caused via ICEO. The scale bar represents 15 µm.

\textbf{Supplementary Movie 8:} Two examples of mixing in the vesicle membrane leading to run and tumble motion of the vesicles at constant AC field conditions. In both examples, the scale bar indicates 20 µm.

\textbf{Supplementary Movie 9:} Large vesicle moving with a smaller vesicle inside, acting as a proof of concept for cargo transport with active vesicles. Only the red channel was used during imaging. The scale bar is 10 µm.

\textbf{Supplementary Movie 10:} Large vesicle containing smaller vesicles bursting upon decreasing the frequency from 10 kHz to 5 kHz as a method for cargo release of the inner small vesicles. The scale bar is 10 µm.

\textbf{Supplementary Movie 11:} Vesicles with alternative morphologies exhibit various dynamical behaviour. Example 1 shows a patchy vesicle causing chiral motion. The second example shows the motion and reconfiguration of a dumbbell-shaped vesicle under different AC field conditions. The scale bars indicate 10 µm in example 1 and 20 µm in example 2.

\newpage

\section*{Supplementary Figures}

\begin{figure}[ht]
\centering
\includegraphics[width=1\textwidth]{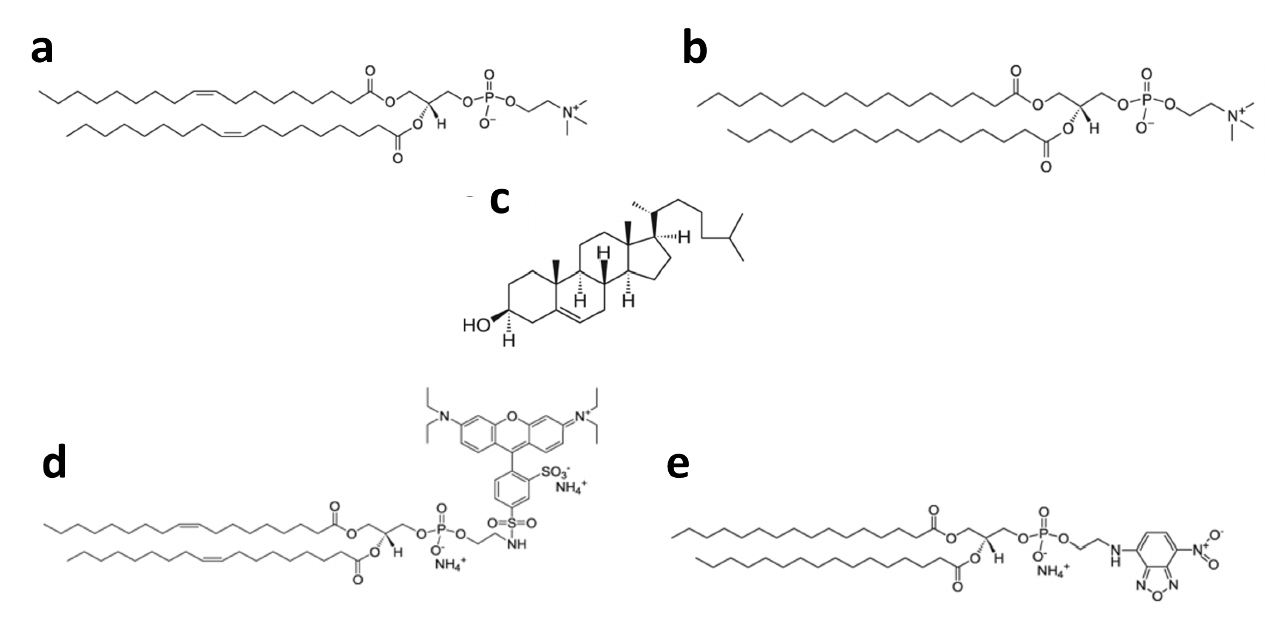}
\caption{Chemical structure of the various phospholipids used to fabricate the Janus GUVs  \textbf{a,} DOPC, \textbf{b,}  DPPC, \textbf{c,} cholesterol, \textbf{d,} RhPE and \textbf{e,} NBDPE.}
\label{fig:S1}
\end{figure}

\begin{figure}[ht]
\centering
\includegraphics[width=1\textwidth]{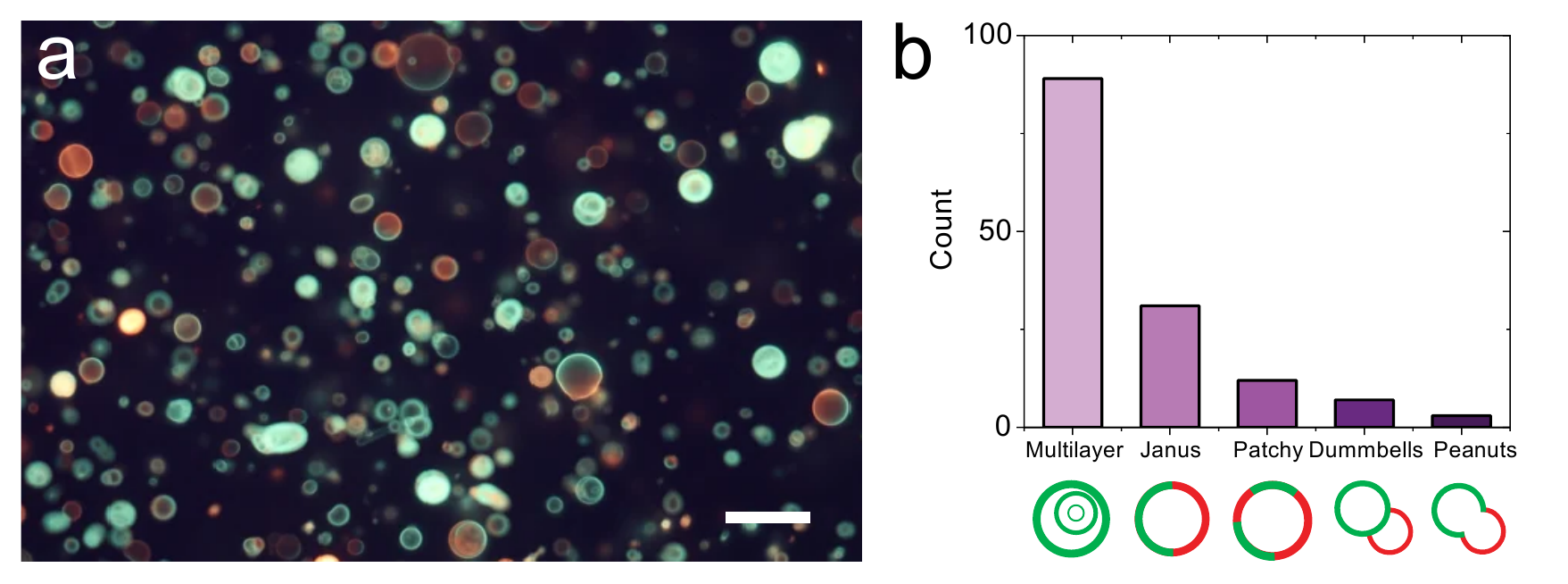}
\caption{\textbf{a,} Fluorescence microscopy picture of a suspension of GUVs produced by electroformation. The scale bar depicts 25 $\mu m$.~\textbf{b,} Population distribution of the various geometries obtained.}
\label{fig:S2}
\end{figure}

\newpage

\begin{figure}[ht]
\centering
\includegraphics[width=1\textwidth]{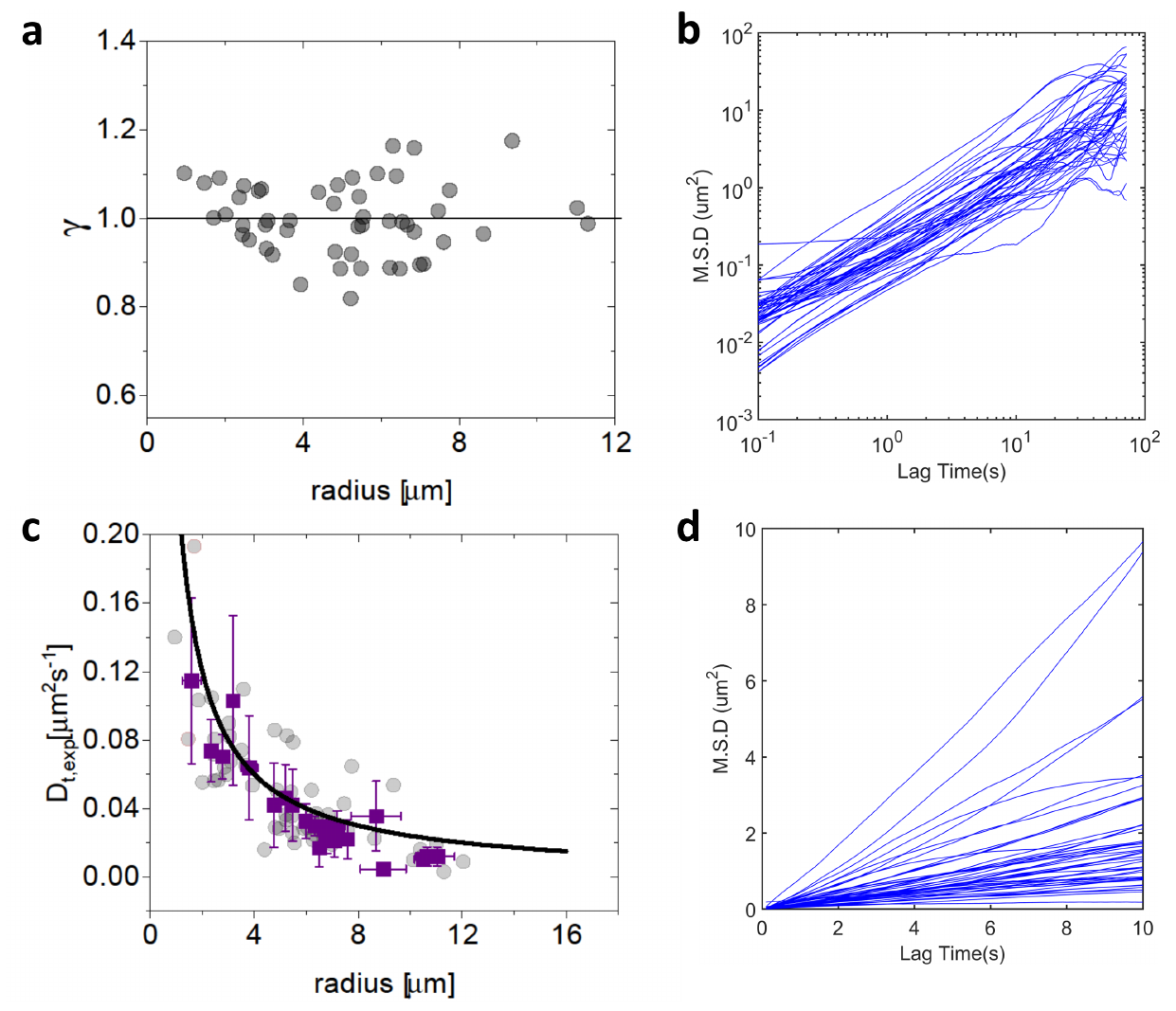}
\caption{\textbf{Passive dynamics of spherical Janus GUVS} \textbf{a,} Values of $\rm \gamma$ as a function of size, obtained by fitting the slope of MSD curves of passive vesicles in a log-log plot. \textbf{b,}  Log-log plot of MSD curves of passive vesicles (0 kHz, 0 V$_{pp}$) used to fit the exponent $\rm \gamma$. \textbf{c,}  Experimental (grey points) and theoretical (black line) translational diffusion coefficients $\rm D_t$ as a function of vesicle size. The purple squares represent the average over three points each with their error bars. The experimental points were obtained by fitting to the MSD curves of passive vesicles and the theoretical curve was calculated using the Stokes-Einstein relation $\rm D_{T}=\frac{k_BT}{6 \pi \eta r}$. \textbf{d,}  MSD curves of passive vesicles (0 kHz, 0 V$_{pp}$), in a linear plot, used to determine the experimental diffusion coefficient $\rm D_t$.}
\label{fig:S3}
\end{figure}

\newpage

\begin{figure}[ht]
\centering
\includegraphics[width=1\textwidth]{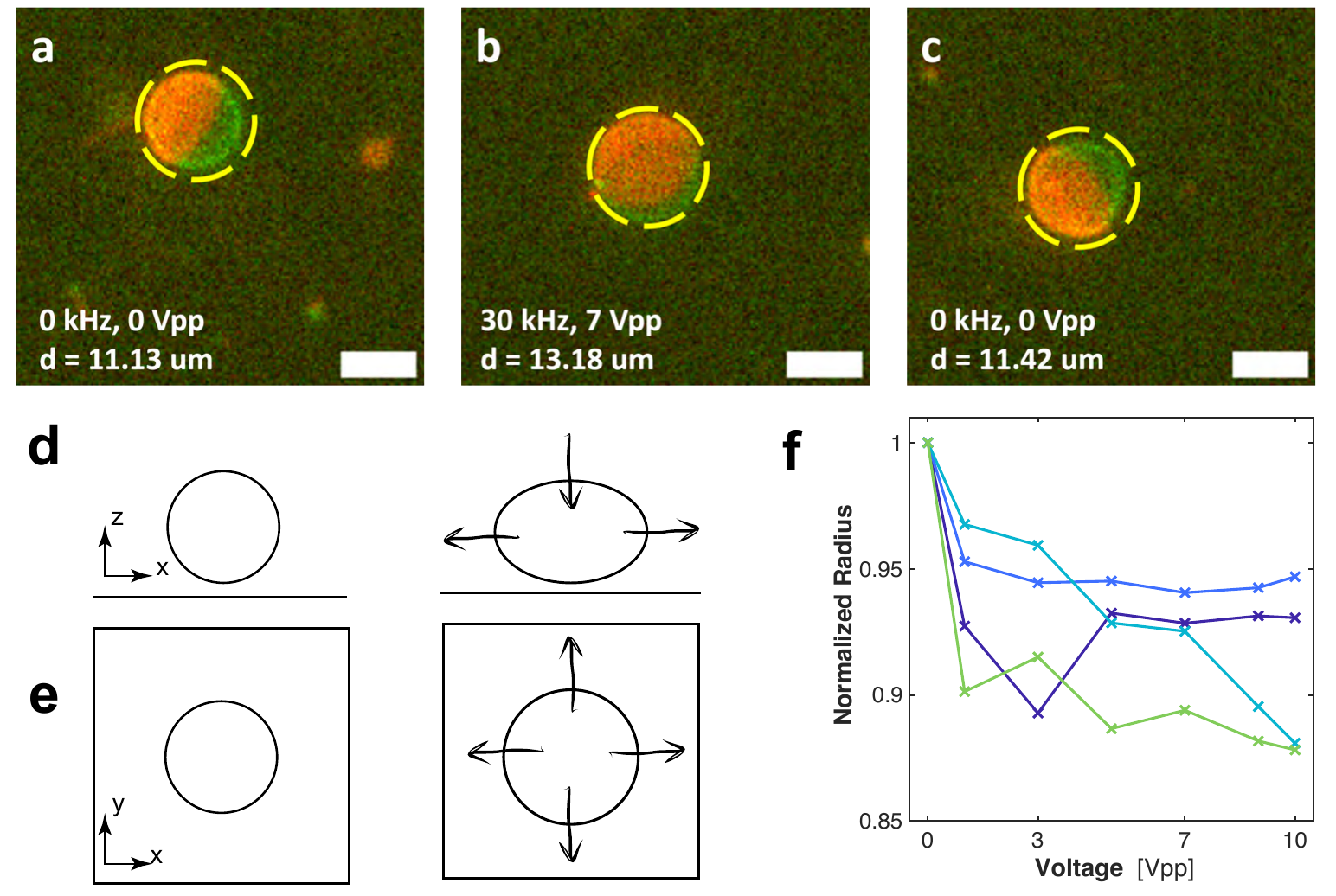}
\caption{\textbf{Membrane deformation under AC fields} \textbf{a,} Vesicle with no applied field. \textbf{b,} Vesicle with a field of 30 kHz and 7 V$_{pp}$ applied. The diameter of the vesicle increased by approximately 2 $\mu$m. \textbf{c,} After switching the field off, the vesicle returned to its original state. All scale bars are 7 $\mu$m and all dashed circles depict a circle of approx.13 $\rm \mu m$. Scheme of the \textbf{d,} side and  \textbf{e,} top view of the vesicle close to the electrode with the field off (left) and on (right). The arrows indicate the direction of the deformation. \textbf{f,} Apparent radius change of individual vesicles as a function of the applied voltage when imaged from the top view. The radius of each particle is normalized to better illustrate the change in radius.}
\label{fig:S4}
\end{figure}

\newpage

\begin{figure}[ht]
\centering
\includegraphics[width=1\textwidth]{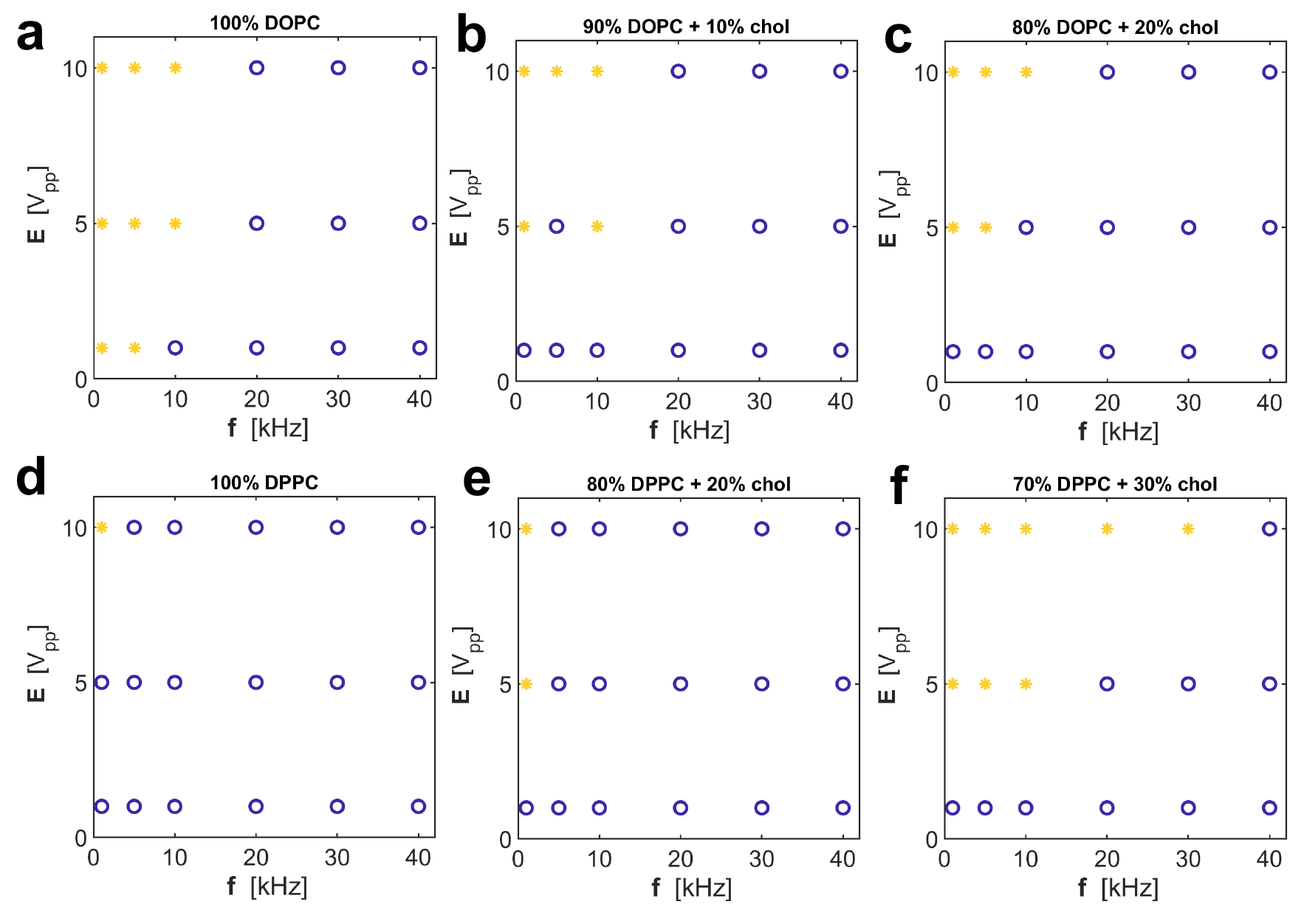}
\caption{\textbf{Dynamical state diagrams of control experiments with non-phase separated vesicles.}  \textbf{a,} 100$\%$ DOPC \textbf{b,}  90$\%$ DOPC + 10$\%$ cholesterol \textbf{c,}  80$\%$ DOPC + 20$\%$ cholesterol \textbf{d,}  100$\%$ DPPC \textbf{e,}  80$\%$ DPPC + 20$\%$ cholesterol \textbf{f,}  70 $\%$ DPPC + 30 $\%$ cholesterol. All DOPC control samples contained 0.1 $\%$ RhPE and all DPPC control samples contained 1$\%$ NBDPE. Legend: \textbf{\textcolor{yellow}{$*$}} breaking of vesicles; \textbf {\textcolor{blue}{$\circ$}} no active motion of vesicles}
\label{fig:S5}
\end{figure}

\newpage

\begin{figure}[ht]
\centering
\includegraphics[width=1\textwidth]{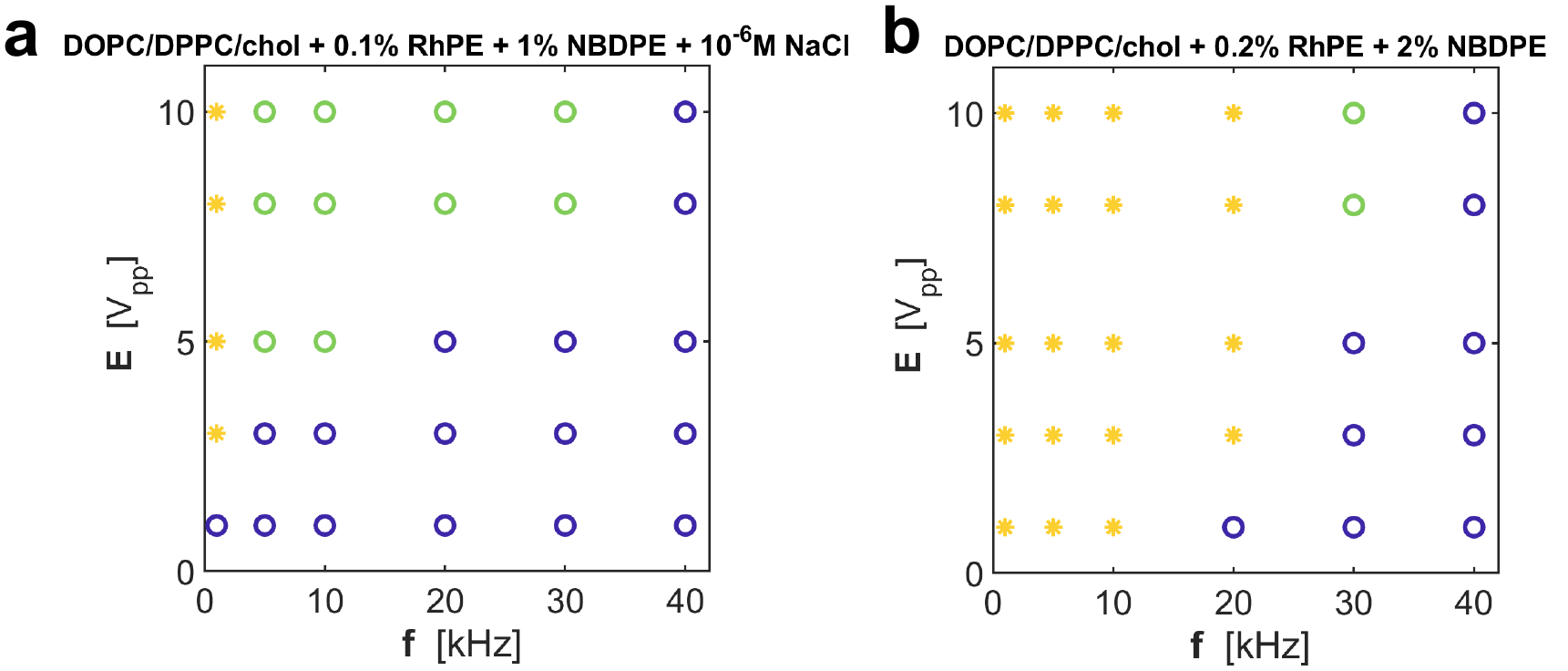}
\caption{\textbf{Dynamical state diagrams of control experiments with phase-separated vesicles.} \textbf{a,} Control experiment with $10^{-6}$ M NaCl in the surrounding solution. \textbf{b,}  Control experiment with doubled fluorophore content (0.2$\%$ RhPE and 2$\%$ NBDPE). Legend: \textcolor{yellow}{\textbf{$*$}} breaking of vesicles; \textbf {\textcolor{green}{$\circ$}} active motion of vesicles \textbf {\textcolor{blue}{$\circ$}} no active motion of vesicles}
\label{fig:S6}
\end{figure}

\newpage

\begin{figure}[ht]
\centering
\includegraphics[width=0.7\textwidth]{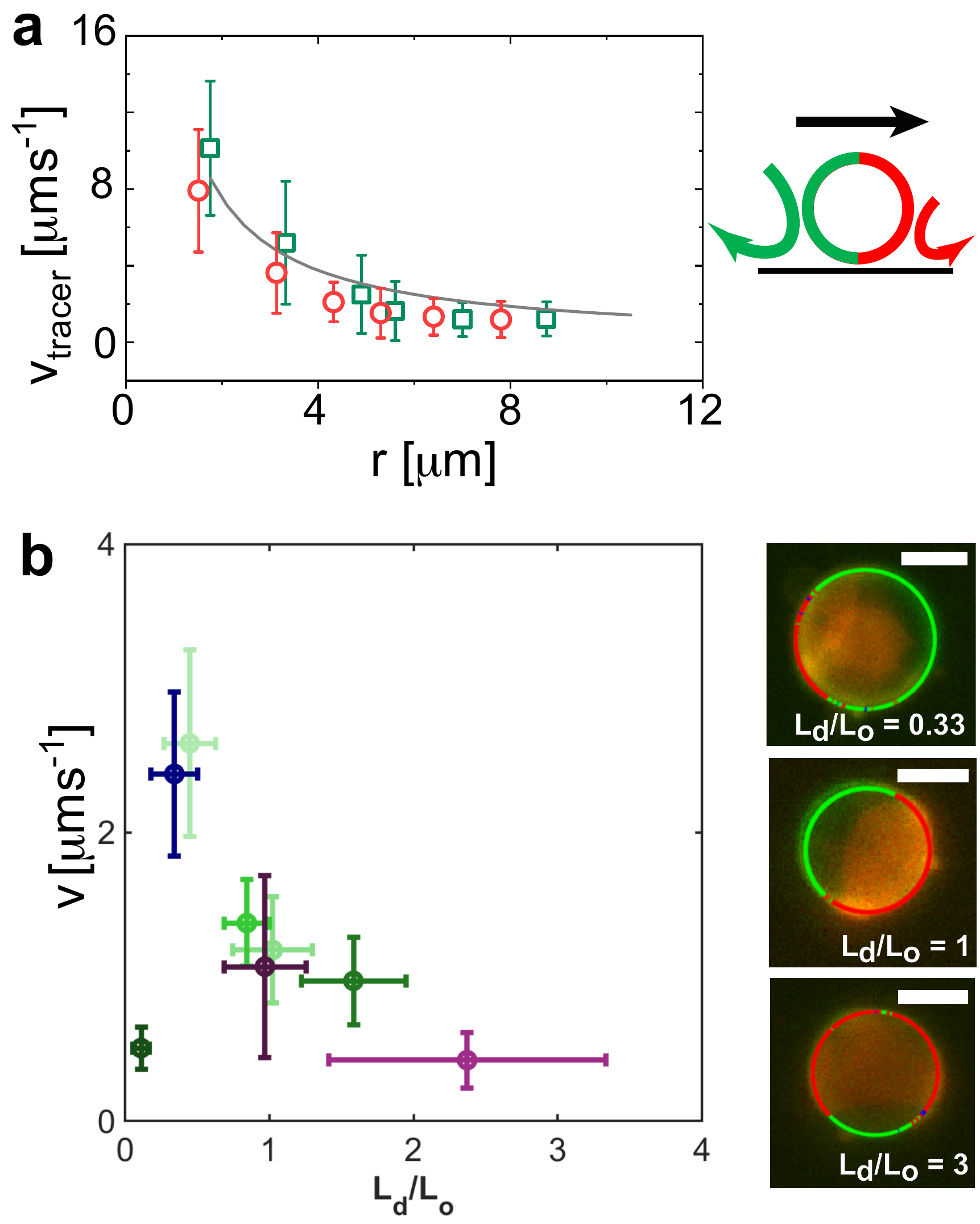}
\caption{\textbf{Asymmetry-dependent GUV swimming speed for Janus configurations} \textbf{a,} Experimental velocity of the tracers $\rm v_{tracer}$ for the $\rm L_o$ (green) and $\rm L_d$ (red) as a function of the distance from the particle ($\rm r$). The solid line represents a decay of $\rm \frac{1}{r}$. The error bars are the standard deviation over more than 50 tracers.\textbf{b,} Velocity at 10 kHz and 9 V$_{pp}$ dependent on the particle asymmetry, defined as the percentage of the liquid-ordered phase ($\rm L_d$, red) in the circumference of the vesicle. The remainder of the circumference is made of the liquid-disordered phase ($\rm L_o$, green). Points in the same colour shade represent vesicles from the same sample and error bars show the standard deviation. (right) Examples of GUVs with different degrees of asymmetry. All scale bars indicate 20 µm.}
\label{fig:S7}
\end{figure}

\newpage

\begin{figure}[ht]
\centering
\includegraphics[width=0.5\textwidth]{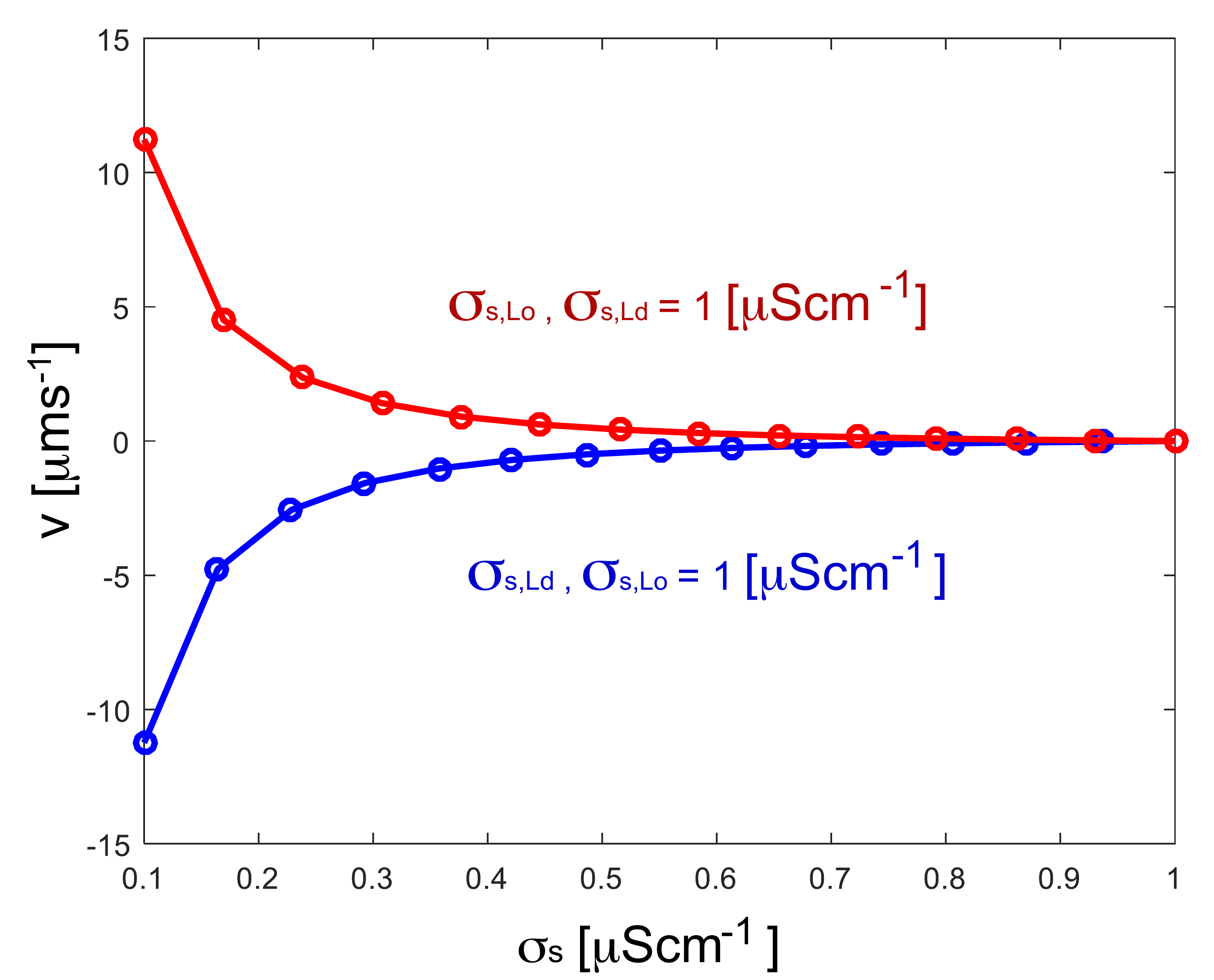}
\caption{\textbf{Velocity dependence on surface conductivity of vesicle hemispheres}, Comsol model numerical simulations results of velocity as a function of the surface conductivity $\rm \sigma_s$ or both hemispheres. The red data correspond to the velocity when the $\rm \sigma_{s,L_{d}}= 1 \mu S cm^{-1}$  and the $\rm \sigma_{s,L_{o}}$ varies, and the blue data when $\rm \sigma_{s,L_{o}}= 1 \mu S cm^{-1}$ and $\rm \sigma_{s,L_{d}}$ varies.}
\label{fig:S8}
\end{figure}

\newpage

\begin{figure}[ht]
\centering
\includegraphics[width=0.9\textwidth]{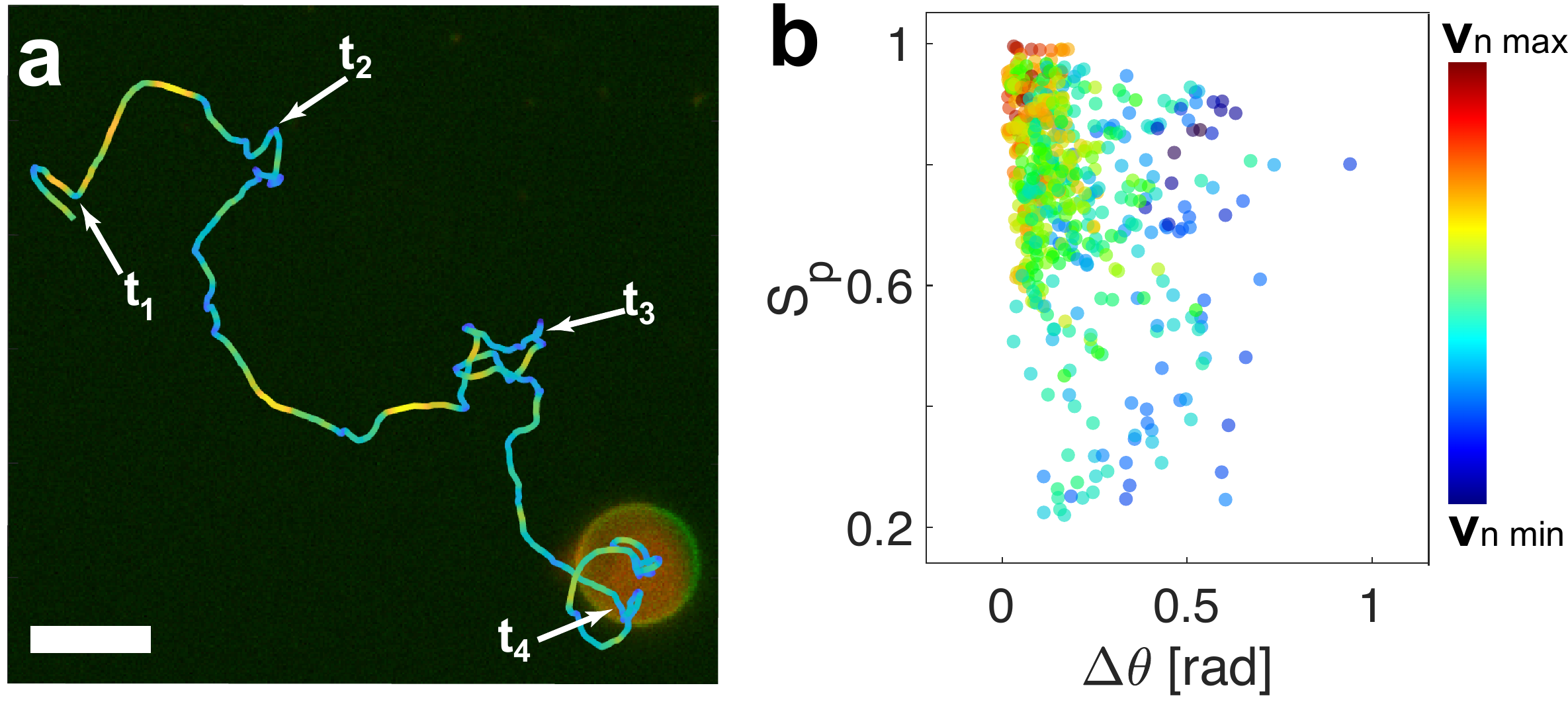}
\caption{\textbf{Run and tumble motion}.\textbf{a,} Trajectory of an active Janus GUV from Fig.4. Scale bar depicts 15 $\rm \mu m$ and \textbf{b,} corresponding characterization of the $\rm S_p$ order parameters as a function of the $\Delta \theta$. The color coding depicts the normalized velocity of the particle.}
\label{fig:S9}
\end{figure}

\newpage

\begin{figure}[ht]
\centering
\includegraphics[width=1\textwidth]{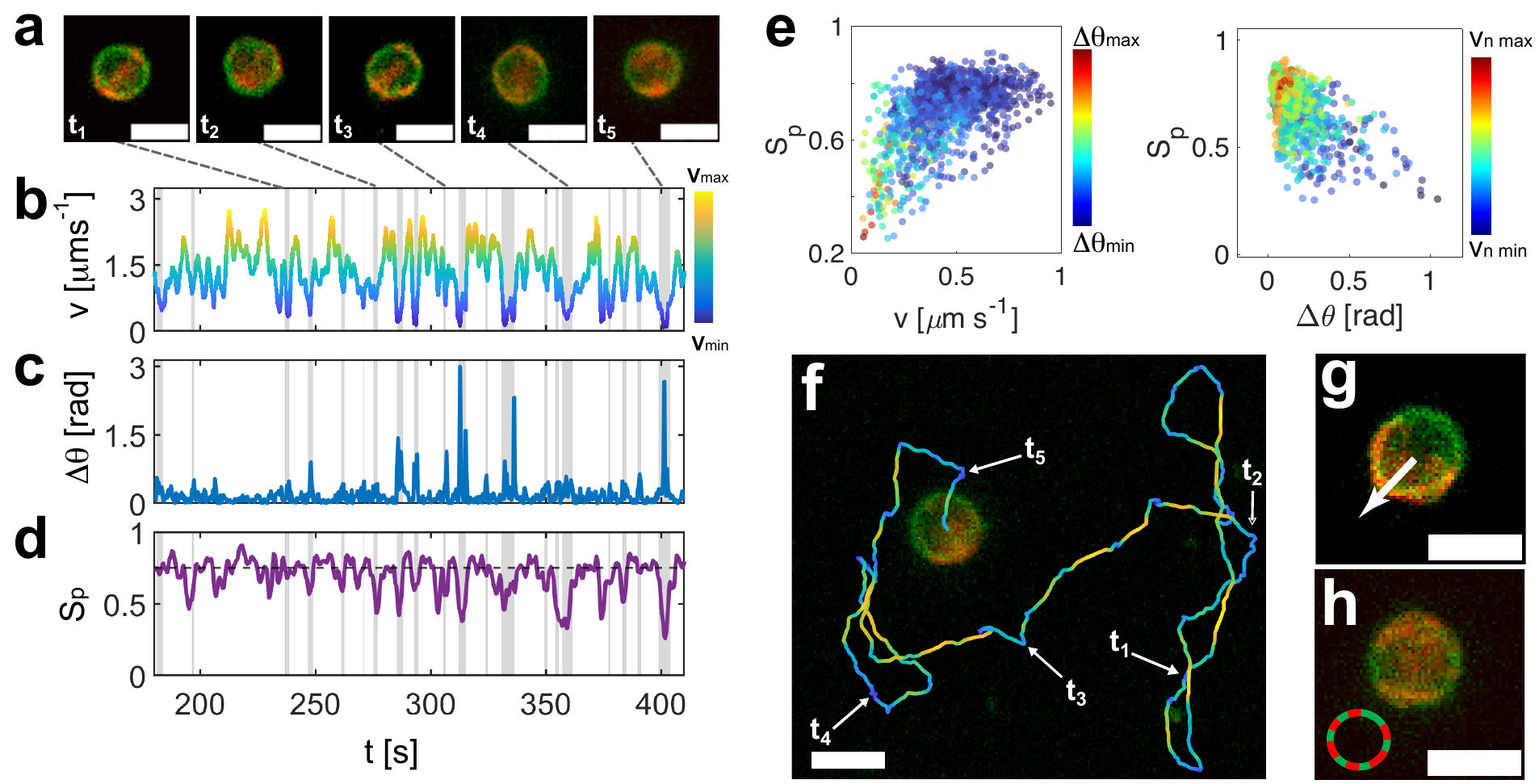}
\caption{\textbf{Analysis of a trajectory depicting run and tumble events at 10 kHz and 9 V$\rm_{pp}$,}\textbf{a,} Snapshots of some identified tumble events. Evolution as a function of time of \textbf{b,} Instantaneous velocity v, \textbf{c,} Variation of Janus vesicle orientation $\rm \Delta \theta$ in between two frames, \textbf{d,} Domain order parameter $\rm S_p$. \textbf{e,} Dependence of the $\rm S_p$ parameter as a function of normalized velocity (left) and $\Delta \theta$ (right).
Fluorescence image using green and red channels overlapping with Janus GUV trajectory, indicating example tumbles from \textbf{a,}. Examples of domain structure for  \textbf{g,} run and \textbf{h,} tumble events. The scale bar depicts 8 $\rm \mu m$.}
\label{fig:S10}
\end{figure}

\newpage

\begin{figure}[ht]
\centering
\includegraphics[width=1\textwidth]{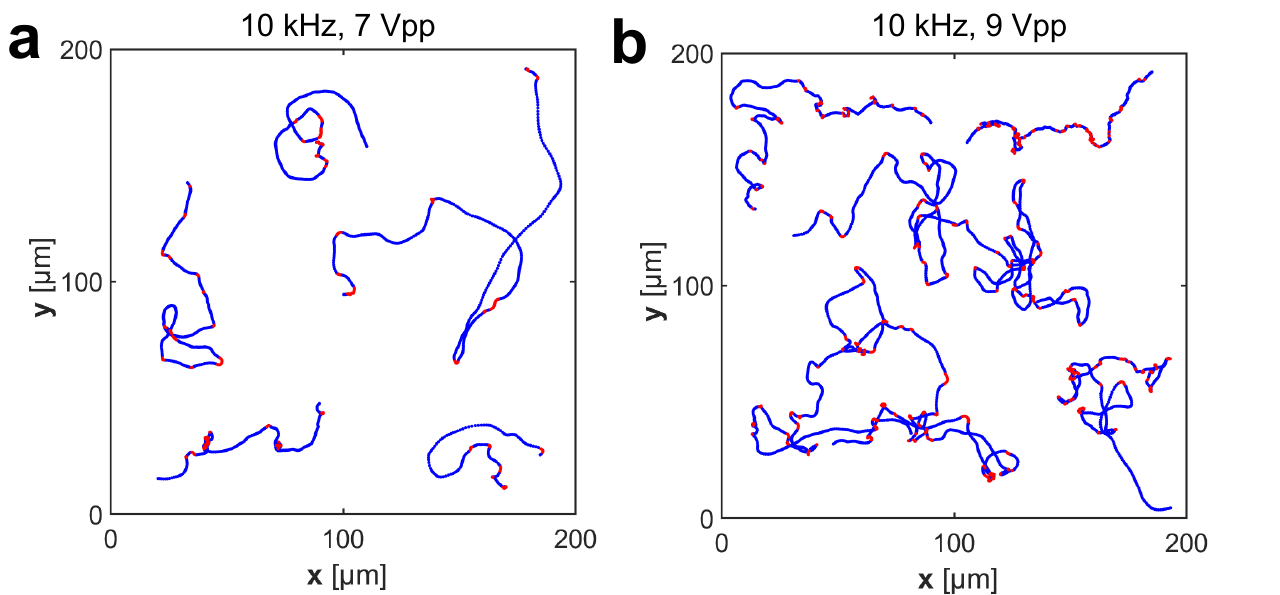}
\caption{\textbf{Run and tumble trajectories}, Example trajectories at \textbf{a,} 10 kHz, 7 V$_{pp}$ and \textbf{b,} 10 kHz, 9 V$_{pp}$ with blue representing runs and red representing the detected tumbles in the trajectories.}
\label{fig:S11}
\end{figure}

\newpage

\begin{figure}[ht]
\centering
\includegraphics[width=1\textwidth]{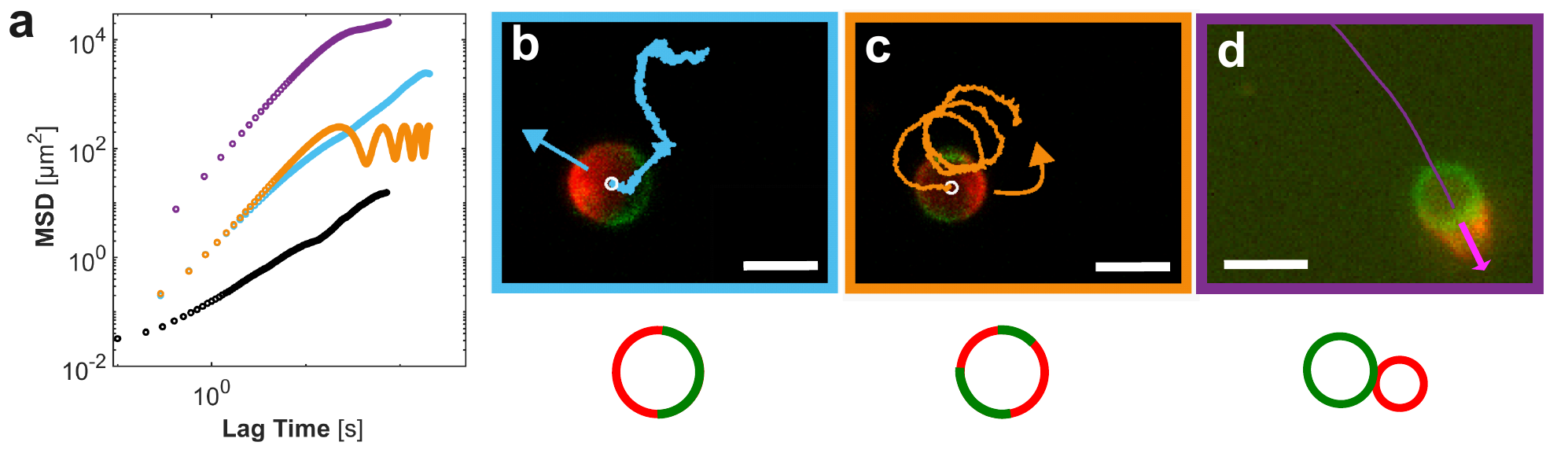}
\caption{\textbf{Dynamics of Janus vesicles with various geometries.} \textbf{a,} MSD curves of vesicles with different morphologies. Blue: Janus vesicle with field on; Black: Janus vesicle with field off; Orange: chiral vesicle; Purple: dumbbell-shaped vesicle. Example trajectories and sketches of \textbf{b,} a Janus vesicle, \textbf{c,} a chiral Vesicle, and \textbf{d,} a dumbbell-shaped vesicle.}
\label{fig:S12}
\end{figure}

\clearpage 

\section*{SUPPLEMENTARY NOTES}

\subsection*{Supplementary Note 1: Calculation of particles electric properties under AC fields} \label{charges}

The time scales of the particle ($\rm \tau_c$) double layer polarization under the effect of an AC electric field have been extensively studied by Squires, Bazant, Ristenpart and Delgado in previous works \cite{Bazant2004,Squires2004,Ristenpart2007}. These time scales can be extracted from the following expressions,

\begin{equation}
\begin{aligned}
\tau_{c} = \frac{\lambda a}{D_{ions}}
%\tau_{c,e} = \frac{\lambda L}{D_{ions}} \\
\end{aligned}
\label{eq:timescales}
\end{equation}

where $\lambda_D$ is the Debye length, a is the radius of the particle, L is the thickness of the experimental cell, and $\rm D_{ions}$ is the diffusion coefficient of the ions in solution. In our experiments, we primarily use MilliQ water, but also carry out control experiments in $10^{-6}$ M of NaCl, so we consider the ionic species Na$^+$ and Cl$^-$ ($\rm D_{ions} = 2x10^{-9} m^2s^{-1}$).

For the phospholipid bilayer, we consider a dielectric shell with $\epsilon_{\rm p}=10$ on average, in reality having a contrast in dielectric constant between the hydrophobic tail and hydrophilic head \cite{waver2003}. The estimation of $\sigma'_{\rm p}$ for the phospholipid bilayer surface was assumed that is analogue to a dielectric surface, in which \cite{CMFactor2017}: 

\begin{equation}
    \sigma_{\rm p} = \sigma_{\rm b} + \frac{2 K_{\rm s}}{r}
\end{equation}
where $\sigma_{\rm b}$ is the bulk conductivity (in Sm$^{-1}$), where for dielectric particles $\sigma_{\rm b} \approx 0$ \cite{Konski1960}, \cite{Ermolina2005}, $K_{\rm s}$ is the surface conductance (in S) of the particle, and r is the particle radius. The value of $K_{\rm s}$ can be approximated as 

\begin{equation}
   K_{\rm s} = \frac{2\sigma_{\rm h}}{\kappa}\left[\frac{D^+}{D^+ + D^-} \left(e^{-\frac{z\zeta_{\rm p} e}{2k_\textrm{B}T}}-1\right)(1+3m^+) + \frac{D^-}{D^+ + D^-} \left(e^{+\frac{z\zeta_{\rm p} e}{2k_\textrm{B}T}}-1\right)(1+3m^-) \right]
\end{equation}
\\
where $\sigma_{\rm h}$ is the medium conductivity, $e$ the electron charge, $z$ the valence of the ions, and $\zeta_{\rm p}$ is the zeta-potential of the particle  ($\rm \zeta_{LO} = -33 mV$, and $\rm \zeta_{LD} = -45 mV $) and $D^\pm$ the diffusion coefficients of the negatively and positively charged ions. The parameters $m^\pm$ are given as \cite{Yang2019} 

\begin{equation}
    m^{\pm} = \frac{2\epsilon_{\rm m}\epsilon_0}{3\eta D^{\pm}} \bigg(\frac{k_{\rm B} T}{ze}\bigg)^2
\end{equation} 
\\
which describe the contribution of electro-osmotic ion flux to $K_{\rm d}$, where $\eta$ is the viscosity of the fluid.
We, therefore, estimate the magnitude of the surface conductivity of each Janus GUV hemisphere composed of a different lipid type. At this salt concentration, the propulsion behaviour of the Janus GUVs is the same as in the experiments with only MilliQ water ($\sigma_{\rm h} \approx 5\times10^{-3}$ Sm$^{-1}$).

Finally, the observed increase in diameter potentially indicates a spreading out from spheres to oblate ellipsoids, as was previously described by Dimova et al. \cite{Aranda2008, Vlahovska2009, Dimova2009}. In our experiments $\rm \sigma_{in} \equiv \sigma_{out}$, with a conductivity of $\rm 1.5*10^{-3}~Sm^{-1}$  (MilliQ water + 25 mM Sucrose) and $\chi=1$, consistent with the prolate deformation. 

\subsection*{Supplementary Note 2: Phase diagram control experiments of Janus GUVs} \label{sec:controldiagram}

We performed a series of control experiments to understand the behaviour of Janus GUVs under the AC field in the current cell configuration.
We first confirm the absence of active motion in the absence of phase separation, and thus, asymmetry. For this, we prepare one phase GUVs for each type of phospholipid with three compositions varying in cholesterol: i) DOPC with 0.1$\%$ RhPE and either 0$\%$, 10$\%$ or 20$\%$ of cholesterol and ii) DPPC with, 1$\%$ NBDPE and either 0$\%$, 20$\%$ or 30$\%$ of cholesterol (Supplementary Fig. \ref{fig:S5}). In these cases, only bursting or passive motion of the vesicle was observed.

Moreover, for the Janus GUVs vesicles, we prepared one control experiment adding  10$^{-6}$ M of NaCl to check the impact of charge screening on the motility of GUVS, and also doubling the amount of fluorophores added to the mixture (Supplementary Fig. \ref{fig:S6}), to relate the amount of fluorophore with the surface charge and thus mobility.

\subsection*{Supplementary Note 3: DEP Force COMSOL simulations}
\label{sec:comsol}

We use the finite-element-based software COMSOL Multiphysics to simulate the direction and velocity of the Janus vesicle. The dielectrophoretic force of a spherical particle in solution under an applied AC electric field is given by \cite{pethig2017dielectrophoresis,Jones1995}

\begin{equation}
    \F_\mathrm{DEP} = \left( \p.\nabla\right)\E
\label{eq:eqFdep1}
\end{equation}
where $\p$ is the electric dipole moment of the particle.  The dielectric dipole moment related to the field by

\begin{equation}
    \textbf{p} = \varepsilon_0\alpha\E
\label{eq:eqp}
\end{equation}
where $\varepsilon_0$ is the free-space electric permittivity and $\alpha$ is the particle polarizability which as units of volume. For a sphere in the quasi-static regime, we have

\begin{equation}
   \rm \alpha = 3 \epsilon_h V_p Re [K(\omega)]
\label{eq:eqalpha}
\end{equation}

where $V_p$ is the particle volume, $\varepsilon_h$  and is the host medium dielectric constant, and $\rm [K(\omega)]$ is the real part of the Clausius Mossoti factor \cite{CMFactor2017}. In this scenario, we use the dielectric protoplast model with a very thin bilipid layer where the external and internal solutions have the same $\epsilon$ and $\sigma$  and $\rm [K(\omega)]$ can be simplified as \cite{Jones1995}

\begin{equation}
\rm [K(\omega)] \approx \frac{c_mR -\epsilon_h}{c_mR+2\epsilon_h},
\label{eq:eqK}
\end{equation}

where $\rm c_m$ is the membrane effective capacitance ($\rm \approx 0.01 Fm^{-2}$), and R is the radius of the vesicle (5 $\rm \mu m $). At the frequency studied here $\rm [K(\omega)] > 0 $. Along with the relations for $\alpha$ and $\p$ in Eq.\ref{eq:eqp} and Eq.\ref{eq:eqalpha}, we arrive at the following expression for the dielectrophoretic force

\begin{equation}
    \F_{DEP} = \varepsilon_0 \alpha \nabla \lVert\E\rVert^2
\end{equation}

We simulate a thin membrane with two hemispheres using the surface conductivity ($\rm \sigma_s$) from the experiments, for a particle size of $\rm 10 \mu m$ diameter. The particle is placed 100 nm from the bottom electrode, and dielectric constant $\rm \epsilon = 10$. The velocity of the active colloidal particle swimming in viscous media at constant speed where  $\rm F_{DEP} = F_{drag}$ can be defined as \cite{CMFactor2017},

\begin{equation}
   v = \frac{\epsilon_0 \alpha \nabla \lVert\E\rVert^2 }{12 \pi \eta a}
\label{eq:vel}
\end{equation}

Thus, from this we obtained that $v = \frac{F_{DEP}}{6\pi \eta R}$.  We can directly calculate the seed velocity of the Janus vesicle as,

\begin{equation}
\rm  \langle \mathbf{v}* \rangle = \frac{3\varepsilon_0 \varepsilon_{m}}{6\pi\eta R}\frac{V_p}{V}\iiint_V \nabla || \textbf{E} ||^2 dV 
\label{eq:eqfVel}
\end{equation}
where $V$ is the simulation domain. From the numerical results, we use the seed horizontal component of the velocity $v_x^* = \langle \mathbf{v}* \rangle.\hat{x}$ and fit it to our experimental data with a fitting prefactor $\beta=\beta'\mathrm{Re}[K(\omega)]$. $\beta'$ accounts for sources of loss in the experimental setup that are not considered in the simulation.

\subsection*{Supplementary Note 4: Detection of run-and-tumble events and analysis} \label{sec:runtumble}

\textbf{Detection of run and tumble}. We detect the running and tumbling events that occur in a trajectory using two criteria, namely that tumbles exhibit large decreases in velocity as well as large changes in the orientation angle. This approach was adapted from previous works by Najafi et al \cite{Najafi2018} and Seyrich et al \cite{Seyrich2018}. We calculate the instantaneous velocity at each step of the trajectories that were previously determined via particle tracking algorithms and then smoothed using the \textit{smooth} function in Matlab to reduce noise. Subsequently, we determine the time $\rm t_{min}$ of the local velocity minima $\rm v(t_{min})$ and the depth  of these local minima as

\begin{equation}
\begin{aligned}
\rm \Delta v = v(t_{max})-v(t_{min})
\end{aligned}
\label{eq: velocity depth}
\end{equation}

where $\rm t_{max}$ is the local velocity maximum closest to $\rm v(t_{min})$.
If $\rm \Delta v$ is larger than a threshold value, in this case, 70$\%$ of $\rm v(t_{min})$, being $\rm \Delta v \geq 0.7v(t_{min})$, then the minimum is counted as a potential tumble. We determine the width of the minima in order to detect the length of the tumble. For this, we use 

\begin{equation}
\begin{aligned}
\rm 0.3 \Delta v + v(t_{min}) \geq v(t)
\end{aligned}
\label{eq: velocity tumble length}
\end{equation}

where the difference in velocities at the minimum $\rm t_{min}$ and the velocity at time $\rm t$ adjacent to the minimum should not be larger than 30$\%$ of $\rm \Delta v$ in order to be part of the tumble. If $\rm t$ is counted as part of the tumble, the next point $\rm t+1$ is evaluated until equation 15 is not fulfilled anymore. We choose and tune the threshold values empirically by comparing the results of the run-and-tumble analysis to videos of some of the vesicles. 

For the second criterion, we consider the change in orientation angle $\rm \Delta \theta$ at each step of the trajectory. As the difference between the angles does not exhibit the same drops as the velocity does and thus it was not possible to find a universal threshold value without receiving many false positives, we use a different approach to determine potential tumbles. We consider the distribution of $\rm \Delta\theta$ which exhibits a narrow high peak containing the small orientation changes of the runs and large tails which represent the large orientation changes of the tumbles. We separate them into two separate Gaussian distributions using the Matlab function \textit{fitgmdist}. The distribution obtained from the tails contains the local maxima of the change in orientation angles $\rm \Delta \theta (t_{max})$ and we consider these points of the trajectories as potential tumbles as well. The length of these tumbles is given by

\begin{equation}
\begin{aligned}
\rm 0.2| \Delta \theta (t_{max})| \leq | \Delta \theta (t)|
\end{aligned}
\label{eq: angle tumble length}
\end{equation}

with which the changes of angles $\rm \Delta\theta$ at times $\rm t$ surrounding the local maxima are included in the tumble if they are larger than 20$\%$ of $\rm \Delta \theta (t_{max})$. 

Finally, we combine both conditions so the parts of the trajectories to which both conditions apply are detected as tumbles, while the remaining parts of the trajectory are classified as runs. \\

\textbf{Order parameter analysis}: For the purpose of analysis, these 2D of the Janus GUVs images are "unwrapped" into a 1D signal representation. The unwrapping process involves tracing around the perimeter of the vesicle in the 2D image and converting this trace into a 1D sequence. This sequence, or signal, maintains the order of red and green labels as they appear around the vesicle's circumference. The unwrapping ensures that the spatial relationships between different regions on the vesicle are preserved in the 1D representation. Once the 1D signal is generated, we precisely identify transitions between regions labelled red and green. These transitions demarcate the boundaries of contiguous segments, allowing us to measure their spatial extent. Such measurements provide quantitative insight into the spatial persistence of each label. Next, we measure the angular distribution by the midpoints of these segments are computed and converted to angular coordinates, with the full circumference of the vesicle equated to ($2\pi$). This transformation facilitates quantitative analysis of the relative positioning of labelled segments around the vesicle. By computing angular differences between adjacent midpoints, we determine the degree of spatial uniformity in label distribution.

\paragraph{Gradient Calculation and identification of transitions}. Given the 1D signal, \( S \), of length \( N \), the gradient is computed to identify the transition points between the red and green labels:

\begin{equation}
    \nabla u_i = u_{i+1} - u_i 
\end{equation}
Considering the periodic boundary conditions:
\begin{equation}
    \nabla u_N = u_1 - u_N.
\end{equation}
The transition from green to red is indicated by a gradient of +1, and from red to green by a gradient of -1. Based on these transitions, we identify the start and end of each step. Next we can compute the length of each step, \( L \), is computed as:
\begin{equation}
    L_i = \text{End}_i - \text{Start}_i,
\end{equation}
where \(\text{End}_i\) and \(\text{Start}_i\) represent the end and start of the \(i^{th}\) step, respectively. 
Once we have the $\text{Start}_i$ and the length of the step $L_i$ we can compute the angular position of each step by first computing the midpoint of each step, and then the position is converted to an angular value considering the full signal:
\begin{equation}
\theta_i = \left( \frac{\text{Midpoint}_i}{N} \right) \times 2\pi
\end{equation}

\paragraph{Signal Fragmentation}.
The gradient of the signal, capturing differences between successive data points, indicates transitions between labelled regions. The sum of the absolute values of this gradient provides a metric for the overall fragmentation of the signal:
\begin{equation}
F = \sum_i \vert \nabla u_i \vert,
\end{equation}

where $F$ denotes the fragmentation measure, and $\nabla u_i $ represents the gradient at the $i^{th}$ position. An increase of $F$ suggests increased fragmentation, indicative of frequent transitions between labelled regions. Note that when the system is fully separated, i.e there are only two regions, the value of $F = 2$ and when the system is at the maximum disorder the value of $F$ would be equal to the number of pixels, $N$, being measured.

\subsection*{Supplementary Note 5: Simulation of Active Brownian Particles} \label{sec:ABP}

We simulate the dynamics of the ABPs by solving the equations of motion in the over-damped limit as \cite{Volpe2014} :

\begin{equation}
\begin{aligned}
x_i&=x_{i-1} + v~cos \varphi_{i-1} \sqrt{2D_T \Delta t} \xi_{x,i} \\
y_i&=y_{i-1} + v~sin \varphi_{i-1} \sqrt{2D_T \Delta t} \xi_{j,i} \\
\varphi_i&=\varphi_{i-1} + \Omega~\Delta \varphi_{i-1} \sqrt{2D_R \Delta t} \xi_{\varphi,i} \\
\end{aligned}
\label{eq:lange}
\end{equation}

where $v$ is the velocity of the particle, $\Omega$ the angular velocity, and $\xi$ represent independent white noise processes. 

We simulate the motion of ABP of the size of our vesicles, with $\rm r = 4.5 \pm 2.1~\mu m$, and average velocity $\rm v = 1.1~\mu m s^{-1}$ from instantaneous velocities differentiating the run and tumbles. Given the absence of torque a priori on our particles, we set the angular velocity $\Omega$ to zero. The temperature is set at 298 K and the viscosity $\eta$ of MilliQ water with 25 mM of sucrose of $\rm 0.001 Pas$. We run simulations for a total number of 50'000 steps, with a $\rm dt$ of 0.4 s between each step, corresponding with the time resolution of our experiments. 

\section*{Supplementary References}

\bibliographystyle{naturemag}
\bibliography{supplementary}